\documentstyle [titlepage,12pt] {report}
\makeatletter

\newcommand{\singlespacing}{\let\CS=\@currsize\renewcommand{\baselinestretch}{1}\tiny\CS}
\newcommand{\oneandahalfspacing}{\let\CS=\@currsize\renewcommand{\baselinestretch}{1.25}\tiny\CS}
\newcommand{\doublespacing}{\let\CS=\@currsize\renewcommand{\baselinestretch}{1.35}\tiny\CS}

\def\@citex[#1]#2{\if@false\immediate\write\@auxout{\string\citation{#2}}\fi
  \def\@citea{}\@cite{\@for\@citeb:=#2\do
    {\@citea\def\@citea{,\linebreak[0]\hskip0pt plus .2em}%
      \@ifundefined{b@\@citeb}%
      {{\bf ?}\@warning{Citation `\@citeb' on page \thepage\space undefined}}%
      \hbox{\csname b@\@citeb\endcsname}}}{#1}}

\newtheorem{rule-def}[theorem]{Rule}

\begin{document}
%\setcounter{chapter}{1}
% defining short form------
\newcommand{\la}{\lambda}
\newcommand{\si}{\sigma}
\newcommand{\ol}{1-\lambda}
\newcommand{\be}{\begin{equation}}
\newcommand{\ee}{\end{equation}}
\newcommand{\bea}{\begin{eqnarray}}
\newcommand{\eea}{\end{eqnarray}}
\newcommand{\nn}{\nonumber}
\newcommand{\lb}{\label}

\begin{center}
{\bf ELECTROMAGNETIC  MASS MODELS IN  GENERAL THEORY OF RELATIVITY:\\
 Ph.D. Thesis}
\end{center}

\vspace{0.5cm}

\begin{center}
{\bf Sumana Bhadra}\\
Sambalpur University, Jyoti Vihar, Burla - 768019, Orissa, India
\end{center}

\vspace{0.25cm}

\begin{center}
Thesis submitted for the degree of \\DOCTOR OF PHILOSOPHY
          IN SCIENCE \\Under the supervision and guidance of\\
 Dr. Saibal Ray \&  Dr. G. Mohanty\\  on January 2007
\end{center}

\vspace{1.0cm}

\begin{center}  
{\bf Abstruct}
\end{center}
``Electromagnetic mass" where gravitational mass and other physical quantities 
originate from the electromagnetic field alone has a century long distinguished 
history. In the introductory chapter we have divided this history into three 
broad categories -- classical, quantum mechanical and general relativistic. 
Each of the categories has been described at a length to get the detailed 
picture of the physical background. Recent developments on Repulsive 
Electromagnetic Mass Models are of special interest in this introductory part 
of the thesis. In this context we have also stated motivation of our work. 
In the subsequent chapters we have presented our results and their physical 
significances. It is concluded that the electromagnetic mass models which are 
the sources of purely electromagnetic origin ``have not only heuristic flavor 
associated with the conjecture of Lorentz but even a physics having 
unconventional yet novel features characterizing their own contributions 
independent of the rest of the physics".

\newpage
\tableofcontents
\pagebreak

\pagestyle{plain}
\pagenumbering{arabic}

\addcontentsline{toc}{chapter}{Synopsis}

 \begin{center}
{\large\bf SYNOPSIS}\\
\end{center}

 The investigations carried out in the thesis
``ELECTROMAGNETIC MASS MODELS IN GENERAL THEORY OF RELATIVITY"
form seven chapters including the introductory and concluding
ones.\\

{\bf Introduction}\\~ The study of ``electromagnetic mass" where
gravitational mass
 and other physical quantities originate from the electromagnetic
 field alone has a century long distinguished history. In the
 introductory chapter we have divided this history into three broad
 categories -- classical, quantum mechanical and general relativistic.
 Each of the categories has been described at a length to get the
detailed picture of the physical background. In the classical part
starting from the Lorentz's Theory of Electrons it includes
Thomson's Concept of Electromagnetic Mass and Abraham's Model for
the Electrons. We have also described the drawbacks of Lorentz's
Model for the Electrons. In this connection Poincar{\'e}'s Theory
of Electrons has been put forward to eliminate the discrepancy in
Lorentz's Model. Einstein's Special Relativistic Model of
Electrons and various other Models including the General
Relativistic one and Quantum Electron Models have been included in
other parts of the history. Recent developments on Repulsive
Electromagnetic Mass Models are of special interest in this
introductory part of the thesis. In this context we have also
stated motivation of our work.\\

{\it CHAPTER II} \\ {\bf Relativistic Electromagnetic Mass Models
with Cosmological Variable $ {\Lambda} $ in Spherically Symmetric
Anisotropic Source}\\~ In the chapter II a class of exact
solutions for the Einstein-Maxwell field equations, which are
obtained  by assuming the erstwhile cosmological constant $
\Lambda $ to be a space-variable scalar, viz., $ \Lambda
=\Lambda(r) $. The source considered here is static, spherically
symmetric and anisotropic charged fluid. The solutions obtained
are matched continuously to the exterior Reissner-Nordstr\"{o}m
solution and each of the four solutions represents an
electromagnetic mass model.\\

{\it CHAPTER III} \\ {\bf Classical Electron Model with Negative
Energy Density in Einstein-Cartan Theory of Gravitation}\\~ The
experimental results regarding the maximum limit of the radius of
the electron $\sim 10^{-16}$ cm and a few of the theoretical works
readily suggest that there might be some negative energy density
regions within the particle as per General Theory of Relativity.
It is argued in the chapter III of the present investigation that
such a negative energy density can also be obtained with a better
physical interpretation in the framework of the Einstein-Cartan
theory.\\

{\it CHAPTER IV} \\ {\bf Energy Density in General Relativity: a
Possible Role for Cosmological Constant}\\~ We consider a static
spherically symmetric charged anisotropic fluid source of radius
$\sim 10^{-16}$ cm by introducing a variable $\Lambda$ dependent
on the radial coordinate $r$ under general relativity in the
chapter IV. From the solution sets a possible role of the
cosmological constant is investigated which indicates the
dependence of energy density on it.\\

{\it CHAPTER V} \\ {\bf Relativistic Electromagnetic Mass Models:
Charged Dust Distribution in Higher Dimensions}\\~ Electromagnetic
mass models are proved to exist in higher dimensional theory of
general relativity corresponding to charged dust distribution. In
the chapter V, along with the general proof, a specific example is
also cited as a supporting candidate.\\

\pagebreak

{\it CHAPTER VI} \\ {\bf Relativistic Anisotropic Charged Fluid
Spheres with Varying Cosmological Constant}\\~ Static spherically
symmetric anisotropic source has been studied for the
Einstein-Maxwell field equations assuming the erstwhile
cosmological constant $ \Lambda $ to be a space-variable scalar,
viz., $ \Lambda = \Lambda(r) $. The solutions thus obtained are
shown to be electromagnetic in origin in the sense that all the
physical parameters including the gravitational mass originate
from the electromagnetic field alone. Moreover, to construct the
models it is also shown that the generally used pure charge
condition, viz., $ \rho + p_r = 0 $ is not always required for
constructing electromagnetic mass models. This is the main theme
of the chapter VI.\\

{\it CHAPTER VII} \\ {\bf Conclusions}\\~ Electromagnetic mass
models which are the sources of purely electromagnetic origin
``have not only heuristic flavor associated with the conjecture of
Lorentz but even a physics having unconventional yet novel
features characterizing their own contributions independent of the
rest of the physics" (Tiwari 2001). This is, as Tiwari (2001)
guess ``may be due to the subtle nature of the mass of the source
(being dependent on the electromagnetic field alone)". Therefore,
in our whole attempt we have tried to explore ``the subtle nature
of the mass of the source". However, to do this under the general
relativistic framework, we have considered Einstein field
equations in its general form, i.e., with cosmological constant
$\Lambda$ which also acts as a source term to the energy-momentum
tensor. If we consider that $\Lambda$ has a variable structure
which is dependent on the radial coordinate of the spherical
distribution, viz., $\Lambda= \Lambda(r)$ then it can be shown
that $\Lambda$ is related to pressure and matter energy density.
Hence it contributes to the effective gravitational mass of the
system.\\ It is seen that equation of state has an important role
in connection to electromagnetic mass model. Therefore, at first
we have obtained electromagnetic mass model under the condition
$\rho+p=0$. However, later on it is shown that electromagnetic
mass model can also be obtained  by using more general condition
$\rho+p \neq 0$.\\ The model considered in our work, in general,
corresponds to a charged sphere with cosmological parameter in
such a way that it does not vanish at the boundary. The idea
behind is that the cosmological parameter is related to the zero
point vacuum energy it should have some finite non-zero value even
at the surface of the bounding system. For this type of spherical
system we can have a class of solutions related to charged as well
as neutral configurations.\\ It can be shown that these models
have positive energy densities everywhere. Their corresponding
radii are always much larger than $10^{-16}$ cm. Furthermore, as
the radii of these models shrink to zero, their total
gravitational mass becomes infinite. It have been shown by Bonnor
and Cooperstock (1989) that an electron must have a negative
energy distribution (at least for some values of the radial
coordinate). In this connection we have shown that the
cosmological parameter $\Lambda$ has a definite role on the energy
density of micro particle, like electron. At an early epoch of the
universe when the numerical value of negative $\Lambda$ was higher
than that of the energy density $\rho$, the later quantity became
a positive one. In the case of decreasing negative value of
$\Lambda$ there was a smooth crossover from positive energy
density to a negative energy density.\\ It is suggested by Bonnor
and Cooperstock (1989) and Herrera and Varela (1994) that spin and
magnetic moment can be introduced to electron through the
Kerr-Newman metric. But, it has been seen that the Kerr-Newman
metric cannot be valid for distance scales of the radius of a
subatomic particle. We, therefore, tackled the problem in the
frame work of Einstein-Cartan theory where torsion and spin are
inherently present. In this case, the only way is to take the spin
to be the `intrinsic angular momentum' that is the spin of quantum
mechanical origin. In our work considering the spins of all the
individual particles are assumed to be oriented along the radial
axis of the spherical systems we have obtained some interesting
solutions with physical validity. \\ Another important point we
would like to mention here that in all the previous investigations
we have studied electromagnetic mass models in 4-dimensional
Einstein-Maxwell space-times only. Therefore, one can ask whether
electromagnetic mass models also can exist in higher dimensional
theory of General Relativity. We have presented a model which
corresponds to spherically symmetric gravitational sources of
purely electromagnetic origin in the space-time of $(n+2)$
dimensional theory of general relativity.\\ We have also taken up
the problem of anisotropic fluid sphere as studied earlier in a
different view point. By expressing $\Lambda$ in terms of electric
field strength $E$ we have explored some possibilities to
construct electromagnetic mass models using the constraint $ \rho
+ p \neq 0 $. We would like to mention here that unlike the
solutions of Gr{\o}n (1986a,b) and Ponce de Leon (1987a,b) in the
present investigation, in general, the electric field (and hence
the cosmological constant) does not vanish at the boundary.
However, it is shown that the class of solutions obtained here are
related to charged as well as neutral systems of Gr{\o}n (1986a,b)
and Ponce de Leon (1987a,b) depending on the values of the
parameter $N$.\\

\pagebreak

\chapter{Introduction}

\hrule
\vspace{0.5cm}
\begin{quote}
 {\it ``...the world stands before us as a great, eternal
 riddle.''}\\
 -- Albert Einstein
\end{quote}
\hrule

\vspace{1.5cm}
 The first elementary particle
was called corpuscle by J. J. Thomson (1881), later named as
electron -- the word proposed by G. Johnstone Stoney in 1891 as a
unit of electronic charge. There were two schools of thought, one
school favoring atomistic world view, and the other believing in
the continuum. The debate continued for several years in absence of
sufficient experimental findings. It is in this respect
that J. J. Thomson's work turns out to be epoch making, and the
credit for the discovery of electron belongs to him. He succeeded
in a determination of the charge to mass ratio and the elementary
charge $e$. The physical world and all its phenomena in principle
can be reduced to the problem of the interaction of the elementary
particles. The four types of fundamental interacting forces are
labeled as electromagnetic, gravitational, strong and weak. Of
these four interactions the electromagnetic one thoroughly
investigated and has been best understood in connection with the
electrically charged particles, especially, the electrons.\\
The
study of electromagnetic mass with a century long distinguished
history can be divided into three broad categories -- classical,
quantum mechanical and general relativistic. Here we shall first
follow the tradition established by the classical electron theory.
Actually, there are two reasons why the classical electron is
studied:\\
a) There is no quantum mechanical model for the
electron;\\
b) It is quite natural to complete the classical
electromagnetic mass theory first before making the transition to
a quantum description of the inertial properties of an electron.\\
In the classical framework Lorentz tried to tackle the problem
of the electrodynamics of moving bodies. To get an overview of
the problem starting from Lorentz, we shall first provide a brief
historical account of the work done by different authors in a
sequence.\\

\section {A Brief Historical Background of the Theories About
the Structure of the Electrons}
\subsection{Lorentz's Theory of Electrons}

 Lorentz's (1892) apparent motivation was to solve the
null result of the Michelson-Morley experiment keeping the existence
of ether as it is. He wanted to represent an electromagnetic world
view in comparison to the Maxwellian electromagnetic theory. Based
on this philosophy he developed his Theory of Electron in 1892.
The basic assumption made by him is that microscopic charged
particles or ions in motion through absolutely resting ether were
the source of the electromagnetic disturbances. He further assumed
that ether permeates the electrical particles and that electrical
particles were perfectly rigid bodies. He considered an electric
field vector $\vec E$ and a magnetic field vector $\vec H$ in this
absolute ether frame and obtained
 \bea
\vec F=\sigma \left(\vec E + \frac{\vec u \times \vec H}{c}\right)
 \eea
where $\sigma \rightarrow$  electric charge density, $u
\rightarrow$ velocity of electric particles and $ c \rightarrow$
velocity of light quanta, photons with velocity $c = 2.99 \times
10^{10} cm/sec$ in vacua.\\
He also considered the Maxwell equations
\be
\vec\nabla.\vec E=4\pi \rho,
\ee
\be
\vec\nabla.\vec B=0,
\ee
\be
\vec\nabla\times\vec E=-\frac{1}{c}\frac{\partial \vec B}{\partial
t}\ee
and
\be
\vec\nabla\times\vec B=\frac{4\pi}{c}\vec J +
\frac{1}{c}\frac{\partial \vec E}{\partial t}
\ee
where $\vec E$,
$\vec B$, $\vec J$ and $c$ respectively are the electric field,
magnetic field, current density and velocity of light in vacua.\\
Now, assuming that the Lorentz-Maxwell equations (1.1-1.5) were
valid in the rest frame of the ether, Lorentz considered the
Galilean transformations
 \bea
\tilde x=x-vt,~ \tilde y=y,~ \tilde z=z,~ \tilde t=t
 \eea
for a reference frame $\tilde S$ moving with an uniform velocity
$v$ in the direction of $x$ with respect to the frame $S$. To retain
the form of the wave
equation describing an electromagnetic disturbance in both the
reference frames $S$ and $\tilde S$ Lorentz introduced new
variables $x'$, $y'$, $z'$ and $t'$ connected with the equations
 \bea
x^{\prime}= \tilde x,~ y^{\prime}=\tilde y,~ z^{\prime}=\tilde z,~
t^{\prime}=t-\frac{v}{c^2}\tilde x
 \eea
(with upto first order of $v/c$). \\
Applying this theory Lorentz was
able to explain the electrodynamics of moving bodies and various
optical phenomena like the propagation of light in dielectrics at
rest as well as in moving media and hence the Fresnel's dragging
effect. The first fruitful idea for explaining the null result of
Michelson-Morley came from Lorentz. He suggested that material
bodies contract when they are moving, and the shortening is only
in the direction of the motion. He proposed that the length of the
interferometer arm parallel to the direction of motion is
shortened by a factor $(1-v^2/2c^2)$. This is known as
Lorentz-Fitz Gerald contraction hypothesis. The more simplified
and generalized theory of Lorentz (1895) on electrodynamics which
involved the concept of local time in contrast to the universal
true time was put forward by him to ensure that the form of
Maxwell equations for charge-free-space remains the same in a
moving frame (at least upto first order in $v/c$). Lorentz assumed
that there are several electrons in each atom which are
elastically bound to an equilibrium position and thus are able to
perform harmonic vibrations with given frequencies. In electric
conductors additional electrons were assumed to move freely. With
these fundamental theoretical tools it was possible to explain a
great number of phenomena -- the absorption, scattering and
refraction of light by matter, the Zeeman effect and many more.
During 1900 to 1903 Lorentz conjectured that a part of the
electron mass might be of electromagnetic origin. However,
for the sake of history we should mention that even before Lorentz there were
other notable scientists, who had expressed the idea of
electromagnetic mass in their works.\\

\subsection{Thomson's Concept of Electromagnetic Mass}

 While studying the interaction of charged
particles Thomson (1881) found that the kinetic energy of a
charged sphere increases by its motion through a medium of finite
specific inductive capacity. He pointed out that the increase in
the kinetic energy was due to the self induced magnetic field of
the charged sphere and calculated the total mass to be \bea m=m_0
+ \mu \eea with $\mu=\alpha e^2/ac^2$ where $m_0$ is the mass of
the charged sphere, $\mu$ is the increased mass, $e$ the electric
charge, $a$ the radius of the sphere and $\alpha$ is a numerical
factor of order unity. Thus he came to the conclusion that ``the
effect of electrification is the same as if the mass of the sphere
were increased ...". This increased mass which clearly is of
electromagnetic origin was termed as {\it electromagnetic mass}.
The works of Thomson on electromagnetic mass were improved upon by
O. Heaviside (1889) showing that the mass of a uniformly moving
charged body varied with velocity. G. F. C. Searle (1897) extended
the work of Heaviside as that the energy of a charged body and
therefore its mass increases with velocity. Later on W. Kaufmann
(1901 a,b,c; 1902 a,b) through a series of experiments on beta
rays established the dependence of the electron mass on velocity.
He (1901b) showed that one-third of the fast moving electron mass
was of electromagnetic origin. Hence it was probably an easy task
for M. Abraham (1902) to speculate that the electron mass was
completely of electromagnetic origin. By reanalyzing his
experimental data Kaufmann (1902b) found that the mass of the
electron is purely of electromagnetic origin.\\

\subsection{Abraham's Model for the Electrons}

 M. Abraham (1902, 1905) proposed the first field theoretical model for the
electron on the basis of Lorentz field equation (1.1) and the
Maxwell field equations (1.2 - 1.5). He assumed the electron to be
a rigid sphere with uniform surface charge distribution. Further,
he considered the electron charge density as a fundamental
quantity. From Kaufmann's experimental work he already knew about
the velocity dependence of electron mass. So, he undertook a
detailed study of the dynamics of electron including a scheme to
derive the electron mass entirely from its self-field. Proposing
the equation of motion for electron as an analogy for Newtonian
equations of motion and considering that the total force acting on
the electron should always vanish, Abraham ultimately got two
types of electron masses, one is the longitudinal mass
($m_{\parallel}$) and the other is transverse mass ($m_{\perp}$)
as \bea m_{\parallel}=\frac{e^2}{2a c^2
\beta^3}\left[\frac{2\beta}{1-\beta^2}\right] -ln
\left(\frac{1+\beta}{1-\beta}\right)\eea
and
\bea
m_{\perp}=\frac{e^2}{4a c^2
\beta^3}\left[(1+\beta^2)ln\left(\frac{1+\beta}{1-\beta}\right) -
2 \beta \right].
 \eea
For low velocities and in the limit $\beta =0$, the electron mass
was
\bea
\mu= m_{\parallel}=
m_{\perp}=\frac{2}{3}\left(\frac{e^2}{a c^2}\right).
\eea
He also
established the electron radius to be of the order of $10^{-13}$
cm. Finally, he claimed that the transverse mass (1.10) was in
agreement with Kaufmann's experimental data to a good
approximation.\\

\subsection{Lorentz's Model for the Electrons}

In 1903 Lorentz extended his electromagnetic theory of 1895. Here
he included the equations of motion of free electron together with
a review of Abraham's model of the electrons. In the further
development of the theory Lorentz speculated that the spherical
electron would experience an ellipsoidal change in its shape while
it is in motion and obtained the transverse mass (the relativistic
mass) for electron as
 \bea
m_{\perp}=\frac{e^2}{6 \pi a c^2 (1-\beta^2)^{1/2}}.
 \eea
 It is then a straight forward way to obtain the relativistic
electromagnetic mass of the field of the electron as
\bea
m^{\prime}_{elec}=\frac{U_{elec}}{c^2}=\frac{1}{2}\left(\frac{e^2}{a
c^2}\right)
 \eea
which is not the same as non-relativistic electromagnetic
mass
\bea
m_{elec}=\frac{2}{3}\left(\frac{e^2}{a c^2}\right).
\eea
Combining these two one can write
\bea
U_{elec}=\frac{3}{4}m_{elec}c^2.
\eea
Certainly this relation unifies gravitation with electromagnetism
and thus through Lorentz the concept of electromagnetic mass was
born. Lorentz then proposed a model of an electron as an extended
body consisting of only pure charge and no matter and the charge
is uniformly distributed on a spherical shell. In his work Lorentz
dealt only with the inertial aspect of mass. Lorentz was certainly
aware of the Newtonian gravitational aspect of mass, but he
probably disregarded gravitational effects because Newtonian
gravitational forces are smaller than electrical forces by
many orders of magnitude.\\

\subsection{Drawbacks of Lorentz's Model}

A discrepancy lies between the two formulas (equations 1.13 and
1.14) for the electromagnetic mass. This problem implies that the
relationship between momentum and velocity for a particle in
Newtonian mechanics differs from that for the completely
electromagnetic electron. This defect can be corrected by merging
this theory with special relativity. The anomaly of factor $4/3$
disappears since it is incompatible with the relativistic
transformation properties. For the finite electron this was first
pointed out by Fermi (1922), but his work did not receive its
recognition till 1965. For point electrons the removal of this
factor was later rediscovered. The inertia and mass of the
classical electron originate from the unbalanced mutual repulsion
of the volume elements of the charge caused by the distorted
electric field of an accelerating electron. However, it is not
clear what keeps the electron stable since the Lorentz's model of
the electron describes its charge as uniformly distributed on a
spherical shell, which means that its volume elements tend to blow
up by repelling one another. This difficulty can be removed by
eliminating the electron structure and assuming the particle to be
a point particle. This indeed produces a new difficulty, i.e.,
when the radius shrinks to zero, the electron's mass becomes
infinitely large. This is the famous self energy problem. It
exists in the classical theory as well as in the quantum theory of
the electron. Its satisfactory solution is not yet known. Thus the
electron is then considered to have a finite extension. The
difficulty with the Lorentz's model was that it had no mechanism
to overcome the electrostatic repulsion of the charge, so that the
body was unstable.\\

\subsection{Poincar{\'e}'s Theory of Electrons}

Poincar{\'e} (1905, 1906), with the aim of overcoming the
instability and inconsistency of Abraham's model with respect to the special
relativistic Lorentz transformations proposed that an attractive
force of cohesive type and consequently non-electromagnetic in
nature can be added so as to just balance the stresses and
establish stability. He first considered the coordinate
transformations in the proper form and called these as Lorentz
transformations which are
 \bea
x^{\prime}= \gamma(x - vt),~ y^{\prime}=y,~
z^{\prime}=z,~t^{\prime}=\gamma(t-vx/c^2).
 \eea
Then he proposed a new Lagrangian which was composed of two parts
 \bea
 L=L_{field}+L_{stress}=-\frac{4}{3}u_0^{em}(1-\beta^2)^{1/2}
\eea
 where $L_{field} \rightarrow$ the
Lagrangian for the electron's self electromagnetic fields,
$L_{stress} \rightarrow$ the Lagrangian related with the so-called
Poincar{\'e} stress and $u_0^{em} \rightarrow$ electromagnetic
energy of the spherical body.\\
This new Lagrangian not only
solved the instability problem but also removed the discrepancy in
the calculation of mass of the electron. In fact, one can show that
in a relativistic theory in which the electron self energy is not
infinite, the self stress will vanish and the particle will be
stable.\\

\subsection{Einstein's Special Relativistic Model of
Electrons}

Einstein's special theory of relativity is based on two unique
postulates, that is, the principle of relativity and the constancy
of velocity of light in vacuum. Using these postulates he could
derive the Lorentz transformations and explain the length
contraction and time dilation as a kinematical consequence of
these transformations. He further showed the covariance of
Maxwell-Lorentz electromagnetic field equations under Lorentz
transformation and that the Lorentz force equation (1.1) is a
consequence of the principle of relativity. The longitudinal and
transverse masses of the electron as given by Einstein (1905) are
\bea m_{\parallel}=\gamma^{3} m_0 \eea and \bea m_{\perp}=\gamma
m_0 \eea where $m_0$ is the rest mass of the electron. These
results are exactly identical to those of Lorentz. Of course, the
dependence of mass on velocity, as given by equation (1.19) does
not mean that only the electromagnetic mass of an electron is a
privileged mass to vary with velocity; rather it is just like any
other kind of mass.\\

\subsection{Various Other Type of Models Including
General Relativistic Electron Models}

 In 1912 G. Mie (1912 a,b) tried to build charged particle
models based on electromagnetic fields alone, so that the mass of
the charged particle like electron could be completely of
electromagnetic origin, and suggested a modification of the
Maxwell-Lorentz field equations. He assumed that the complete
electromagnetic field is determined by ten universal quantities
which are functions of the four-potentials $A^i$ and the Maxwell
tensors $F^{ij}$. But this unitary field theory ultimately failed.
The other workers on the theory of unitary field, with different
view points, were M. Born and L. Infeld (1934), B. Hoffmann (1935
a,b) and F. Bopp (1940, 1943). The search for a solution to the
problem of the electron structure were made by H. Weyl (1918a,b;
1919), T. Kaluza (1921) and O. Klein (1926a,b; 1928) in the realm
of unified field theory where gravitational and electromagnetic
fields have been unified into a single theory. In 1919, the first
general relativistic approach towards an electromagnetic mass was
 put forward by Einstein. To overcome the drawbacks of Mie's theory
Einstein proposed a model where gravitational forces would provide
the necessary stability to the electron and the contribution
to the mass would also come from it. The discovery of quantum mechanics
in 1925 -- 1926 called for an extension of the classical theory of
the electron to the atomic and subatomic domain. A very important
and unexpected property of electrons was found almost
simultaneously with the establishment of this new mechanics. It
was discovered that the electron has an intrinsic angular
momentum, a spin and a magnetic moment associated with it.
Essential progress towards a relativistic quantum mechanical
description of electrons was made by Dirac (1928). It was Dirac
who was able to devise a wave equation for the electron which
fulfilled the relativistic requirements. Actually quantum
mechanics treats electron as a point-like charged particle with
spin and hence extended electron could not be accommodated within
it. Thus it seems that for a better description of the electron structure
with a spin in the General Relativistic framwork it is already seen that
the Einstein-Cartan-Maxwell or Einstein-Maxwell-Dirac space-times
is preferable rather than the Einstein-Maxwell space-times
(Ray and Bhadra 2004a). In the various classical
point charge theories (Mehra 1973) the electron is treated as a point charge
having a pure charge without any structure. But these theories cannot
overcome the self energy problem of the electron which becomes
infinite at its location. This infinite self energy problem can be
solved by considering an extended charge distribution for the
electron. But these theories have no satisfactory quantum
versions. In the framework of Einstein's general theory of
relativity (which was proposed in 1916) a lot of work has been
carried out by different authors on charged body. Some models
which were developed to study the structure of electron are due to
Kyle and Martin (1967), Cohen and Cohen (1969) and Baylin and Eimerl
(1972). Katz and Horwitz (1971) and L\'{o}pez (1984) have
developed classical extended electron models from the general
relativistic point of view. In all these models an electron has
been considered to be a microscopic sphere of charged perfect fluid
or as a spherical shell of matter embodied by charge.\\

\section{A Short Account of the Recent Developments in the
Electromagnetic Mass Models}

So long we have seen in the theories that in the Special
Relativity and even in the General Relativity the mass of the electron
has been considered to contain two parts, that is, the
non-electromagnetic part and the electromagnetic part.
The exceptions are Lorentz's and Abraham's theories which have
independently considered the mass of electron to be completely of
electromagnetic origin. In the following section we shall give a brief
description of the above idea showing how the conjectures of
Lorentz-Abraham may be revived on the ground of the General
Relativity.

\subsection{Repulsive Electromagnetic Mass Models: Electron Type}

 Repulsive gravitation is produced by the negative mass
of the polarized vacuum. The vacuum fluid obeying an equation of
state $\rho=-p$ was taken by most of the workers for the
construction of electromagnetic mass model. By considering the
relation between the metric coefficients i.e., $g_{00}g_{11} =-1$
(which for both the Schwarzschild and the Reissner-Nordstr\"{o}m
matrices equivalent to the relation $\nu + \lambda=0$) to be valid
inside a charged perfect fluid distribution, it is shown by Tiwari
et al. (1984) that the mass energy density and the pressure of the
distribution are of electromagnetic origin. In the absence of
charge, however, there exists no interior solutions. A particular
solution which confirms the same and matches smoothly with the
exterior Reissner-Nordstr\"{o}m was obtained by them. This
solution represents a charged particle whose mass is entirely of
electromagnetic origin. The pressure being negative here the model
is under tension and hence the source is of repulsive nature. In
the approach taken by Gautreau (1985), following Tiwari et al.
(1984), the electron's mass is associated with the Schwarzschild
gravitational mass given by general relativity and not with the
inertial mass used by Lorentz (1904). In this case the
Schwarzschild mass of an extended charged body as seen at infinity
arises from the charge as well as the matter possessed by it. Here
the field equations for a Lorentz type pure charge extended
electron are obtained by setting the matter terms equal to zero in
the field equations for a spherically symmetric charged perfect
fluid. An explicit solution to the pure charge field equations are
examined by Gautreau (1985). L\'{o}pez (1984) proposed a classical
model of the spinning electron in which the particle is the source
of the Kerr-Newman field. The electron here is regarded as a
charged rotating shell with surface tension. The phenomenon of
repulsive vacuum gravitation proved to be of importance in
cosmology with appearance of the inflationary universe models.
Gr{\o}n (1985) pointed out the possibility that repulsive
gravitation may be of importance also in connection with
elementary particle models. This possibility was realized by
Tiwari et al. (1984) and by L\'{o}pez (1984). Poincar{\'e}
stresses were explained by them as being due to vacuum
polarization in connection with a recently presented class of
electromagnetic mass models in general relativity. The
gravitational blue shift of light is explained as being due to
repulsive gravitation produced by the negative gravitational mass
of the polarized vacuum. Gr{\o}n (1985) pointed out that the
electron model of L\'{o}pez (1984), which includes spin, and which
is a source of the Kerr-Newman field gives rise to repulsive
gravitation.\\ Assuming an implicit relation among the unknown
physical parameters viz., the pressure `$p$, the charge density
$\sigma$ and the electromagnetic potential $\phi$, it has been
shown by Tiwari et al. (1986) that $\phi$ satisfies the well known
Lane-Emden equation. Electromagnetic mass models corresponding to
the exact solutions of the Lane-Emden equation were obtained by
them. The radii of some of the models were compared by them with
the ``classical electron radius''. L\'{o}pez (1986) analyzed the
stability of a classical ellipsoidal electron model. The model was
found to be stable under oscillations which change the size of the
ellipsoid without altering its shape. It is further shown by
L\'{o}pez (1986) that angular momentum conservation does not allow
the existence of other oscillation models. Ponce de Leon (1987a)
investigated the relation $g_{00}g_{11} =-1$ in the case in which
the interior is filled with imperfect fluid. He found that the
core of such a distribution is gravitationally repulsive provided
the energy density is positive. Ponce de Leon (1988) also
investigated the different aspect of the phenomenon of
gravitational repulsion in static sources of the
Reissner-Nordstr\"{o}m field. He found that in the case of perfect
fluid spheres there exists a close relation between the
gravitational repulsion and the Weyl curvature tensor. Ponce de
Leon (1988) proved that the static source of the
Reissner-Nordstr\"{o}m field gives rise to gravitational repulsion
only if the pure gravitational field energy inside the sphere is
negative. It is also proved that although the gravitational
repulsion always takes place in the interior of a charged perfect
fluid sphere when its radius is less than the classical electron
radius, this is not necessarily so either in the case of
anisotropic charged spheres or if the net charge of the body is
concentrated at its boundary only. He further found that the
charge contributes negatively to the effective gravitational mass,
in the sense that an increase in the charge causes a decrease in
gravitational mass. He explained the gravitational repulsion as
being due to this negative contribution rather than the strain of
vacuum because of vacuum polarization. Bonnor and Cooperstock
(1989) found, by modelling the electron as a charged sphere
obeying Einstein-Maxwell theory, that it must contain some
negative rest mass. The total gravitational mass within this
sphere is negative which is one of the assumptions made in
singularity theorems of general relativity. L{\'o}pez (1992)
constructed a classical model of the spinning electron in general
relativity consisting of a rotating charge distribution with
Poincar{\'e} stresses. Obviously he obtained a class of interior
solutions of the Kerr-Newman field. The negative pressures or
tensions obtained here are identified with the cohesive forces
introduced by Poincar{\'e} (1905, 1906) to stabilize the Lorentz
electron model. They are shown by L\'{o}pez (1992) to be the
source of a negative gravitational mass density and thereby of the
violation of the energy conditions inside the electrons. Herrera
and Varela (1994) pointed out the role played by the negative rest
mass as mentioned in the work of Bonnor and Cooperstock (1989).
Here the electron is modeled as a spherically symmetric charged
distribution of matter deprived of spin and magnetic moment. Since
the electrostatic energy of a point charge is infinite, the only
way to produce a finite total mass is the presence of an infinite
amount of negative energy at the center of symmetry. They (Herrera
and Varela 1994), by analyzing some extended electron models,
showed that negative energy distributions result from the
requirement that the total mass of these models remains constant
in the limit of a point particle. Tiwari and Ray (1996) dealt with
a model which is the charged generalization of static dust sphere
in Einstein-Cartan theory. They obtained a set of solutions with
torsion and spin which represents an electromagnetic mass model.
Blinder (2001) proposed a model for the classical electron as a
point charge with finite electromagnetic self energy. Modified
form of the Reissner-Nordstr\"{o}m and Kerr-Newman solutions of
the electromagnetic equations were derived. Moreover, the self
interaction of a charged particles with its own electromagnetic
field was shown to be equivalent to its reaction to the vacuum
polarization.

\subsection{Repulsive Electromagnetic Mass Models: Stellar Type}

The work of Ray and Das (2002) which is concerned with the charged
analogue of Bayin's work (1978) related to Tolman's type and
presents astrophysically interesting aspects of stellar structure.
However, in a static spherically symmetric Einstein-Maxwell
space-time this class of astrophysical solution
 found out by Pant and Sah (1979) and Ray and Das (2002) has
been revisited in connection with the phenomenological relationship
between the gravitational and electromagnetic fields (Ray and Das
2004). Considering Riccati equation with known value of charge $q$
for the total charge on the sphere in the following form
\begin{eqnarray}
q(a) = Ka^n
\end{eqnarray}
they have shown in one of the cases that the gravitational mass
for $n=1$ can be given as
\begin{eqnarray}
m = q^2 + a_0a_1\left(\frac{q}{K}\right)^2 +
{a_1}^2\left(\frac{q}{K}\right)^3.
\end{eqnarray}
 It is thus qualitatively shown that the charged relativistic stars
of Tolman (1939) and Bayin (1978) type are of purely
electromagnetic origin. Obviously, the existence of this type of
astrophysical solutions is a probable support to the extension of Lorentz's
conjecture that electron-like extended charged particle possesses
only `electromagnetic mass' and no `material mass'.

In this connection some known static charged fluid spheres of
Tolman-VI type solutions have been reexamined and the
gravitational masses are shown to be of electromagnetic origin by
Ray and Das (2007a). They have considered a more general form of
the gravitational mass as follows:
\begin{equation}
m = \frac{n(2 - n)a^2 + 2q^2}{2(1 + 2n - n^2)a}
\end{equation}
where for physical viability the values on $n$ to be assigned are
$0 \leq n \leq 2$.

For the specific choice $K = 1/\sqrt 2$ and $n = 1$ of these
parameters, the {\it ansatz} expressed in equation (1.20) reduces to
$q(a)/a = 1/\sqrt{2}$, where $a$ is the radius of the sphere. It
is interesting to note that for this charge-radius ratio all the
perfect fluid equations of state reduce to the form $\rho + p = 0$
which is known as the `pure charge condition' (Gautreau 1985) and
also imperfect-fluid equation of state in the literature for the
matter distribution under consideration is in tension and hence
the matter is named as a `false vacuum' or `degenerate vacuum' or
`$\rho$-vacuum' (Davies 1984; Blome and Priester 1984; Hogan 1984;
Kaiser and Stebbins 1984).

Ray and Das (2007b) have again considered the Einstein-Maxwell
space-time in connection with some of the astrophysical solutions
previously obtained by Tolman (1939) and Bayin (1978). The effect
of charge inclusion in these solutions has been investigated
thoroughly and the nature of fluid pressure and mass density
throughout the sphere have also been discussed. Mass-radius and
mass-charge relations have been found out for various cases of the
charged matter distribution. Two cases are obtained where perfect
fluid with positive pressures gives rise to {\em electromagnetic
mass} models such that gravitational mass is of purely
electromagnetic origin. The stability conditions have been
investigated for all these Tolman-Bayin type static charged
perfect fluid solutions in connection with the stellar
configurations.

\subsection{Lorentz's Electromagnetic Mass: a Clue
for Unification?}

 Ray (2007) in his review of the electromagnetic mass model by
Lorentz has described the philosophical perspectives and
given a historical account of this idea, especially, in the light of
Einstein's Special Relativistic formula ${E = mc^2}$. It is known
that, at distances below $10^{-32}$ m, the strong, weak and
electromagnetic interactions are ``different facets of one
universal interaction''(Georgy, Quinn and Weinberg 1974, Wilczek
1998). This is already confirmed by (i) the theories of the
unification of electricity and magnetism by Maxwell, (ii) that of
earth's gravity and universal gravitation by Newton and (iii)
``...the unified weak and electromagnetic interaction between
elementary particles..." by S. Glashow, A. Salam and S. Weinberg
for which Nobel Prize was awarded to them in 1979 . Therefore,
as regards unification scheme, Ray (2004) has argued that though
there has been much progress towards a unification of all the
other forces -- strong, electromagnetic and weak -- in the Grand Unified
Theory (GUT), gravity has not been included in the scheme. In this
context it has also been mentioned that there are some problems
with gravity: (i) the strength of the gravitational interaction
is enormously weaker than any other force (the hierarchy problem)
and (ii) General Theory of Relativity does not consider gravity a force,
rather a kind of field for which a body rolls down
along the space-time curvature (the field theoretical problem). As
a probable alternative solution to this problem  Ray (2004) has
put forward Lorentz's conjecture of `electromagnetic mass' and
suggested that this may be a competent candidate of the long desired
unification.

 \section{Motivation and
Discussion of Our Investigation}

Our motivation to work on relativistic electromagnetic
mass models is based on a slightly different approach. We have
studied the role of the cosmological constant in constructing
the electromagnetic mass model. This is the central part of our work.
It has been observed that Tiwari et al. (1984) and other authors
constructed electromagnetic mass models without considering any
$\Lambda$ term. So, we have considered here $\Lambda$ in the
Einstein field equations by assuming it to be a scalar rather than
a constant and hence re-examined the work of Ray and Ray (1993)
and Tiwari and Ray (1996). We obtained a class of exact solutions
for the Einstein-Maxwell field equations by assuming
the cosmological constant to be a space variable scalar, i.e.,
$\Lambda=\Lambda(r)$. The source considered in the chapter II is
static, spherically symmetric and anisotropic charged fluid type.
The solutions obtained are matched continuously to the exterior
Reissner-Nordstr\"{o}m solution and each of the four solutions
represents an electromagnetic mass model. \\
In chapter III we have examined whether even without employing
the vacuum fluid equation of
state $\rho + p =0$ a stable model with electromagnetic mass can
be constructed. Here we have considered a charged anisotropic
static spherically symmetric fluid source of finite radius. The
field equations thus obtained under certain mathematical
assumptions yield a set of solutions which are shown to be
electromagnetic in origin. Electromagnetic mass models have been
studied by several authors under the special assumption $\rho + p
=0$. Here we have shown that even for $\rho + p \neq 0$
electromagnetic mass model can be constructed. This is one of the
motivations of the present investigation. However, the same
question whether there exists any electromagnetic mass models
where this condition $\rho + p \neq 0$ is violated was addressed
by Tiwari et al. (1991) and obtained electromagnetic mass model in
the isotropic and axially symmetric matter distribution or charged
dust case only, whereas in chapter IV we have searched a solution
by employing a relation between the radial and tangential
pressures as $p_{\perp}=p_r+\alpha q^2 r^2$. Our aim is to see if
there is any effect of space dependent $\Lambda$ on the energy
density of classical electron. Here we have considered an extended
static spherically symmetric distribution of an elementary
particle like electron having the radius of the order of
$10^{-16}$ cm. It is already suggested that there might be some
negative energy density regions within the particle in the general
theory of relativity (Bonnor and Cooperstock 1989, Herrera and
Varela 1994). It is, therefore, argued in the present
investigation that such a negative energy density also can be
obtained with a better physical interpretation in the framework
of Einstein-Cartan theory. \\
In many theories higher dimensions
play an important role, specially in superstring theory which
demands more than usual four dimensional space time. This is also
true in studying the models regarding unification of gravitational
force with other fundamental forces in nature. So long
electromagnetic mass model has been studied extensively in the
four dimensional space-times of the General Relativity. Here, in
chapter VI, we have presented electromagnetic mass model in the
space-time of higher dimensional theory of general relativity.
Under this motivation we have considered here a static spherically
symmetric charged dust distribution corresponding to higher
dimensional theory of general relativity. \\
  In chapter VI we have studied static spherically symmetric
 anisotropic source for the Einstein-Maxwell space-times assuming the
erstwhile cosmological constant $ \Lambda $ dependent on the
spatial coordinate, viz., $ \Lambda = \Lambda(r) $. It is shown
that the solutions thus obtained are of electromagnetic in origin
in the sense that all the physical parameters including the
gravitational mass originate from the electromagnetic field alone.
It is also shown that the generally used pure charge condition,
viz., $ \rho + p_r = 0 $ is not always required for constructing
electromagnetic mass models.\\
The concluding chapter VII offers a general discussion on the whole work
along with the future scope of the field of electromagnetic mass models
in General Relativity.

\pagebreak

\chapter{Relativistic Electromagnetic Mass Models with
Cosmological Variable $ {\Lambda} $ in Spherically Symmetric
Anisotropic Source}

\vspace{1.0cm}

\hrule \vspace{0.5cm}
 \begin{quote}
 {\it ``Whether one or the other of these methods will lead to\\
the anticipated ``world law'' must be left to future research.''}\\
-- Max Born (1962)
\end{quote}
\hrule

\vspace{1.5cm}
 \section{Introduction}

 A very important problem in cosmology is that of the cosmological constant the
present value of which is infinitesimally small $(\Lambda
\le10^{-56} cm^{-2})$. However, it is believed that the smallness
of the value of $ \Lambda $ at the present epoch  is because of
the Universe being so very old (Beesham 1993). This suggests that
the $\Lambda$ can not be a constant. It will rather be a variable,
dependent on coordinates -- either on space or on time or on both
(Sakharov 1968; Gunn and Tinslay 1975;  Lau 1985; Bertolami
1986a,b; {\"O}zer and Taha 1986; Reuter and Wetterich 1987; Freese
et al. 1987; Peebles and Ratra 1988; Wampler and Burke 1988; Ratra
and Peebles 1988; Weinberg 1989; Berman et al. 1989; Chen and Wu
1990; Berman and Som 1990; Abdel-Rahman 1990; Berman 1990a,b;
Berman 1991a,b; Sistero 1991; Kalligas et al. 1992; Carvalho et
al. 1992; Ng 1992; Beesham 1993 and Tiwari and Ray 1996). \\
 Now, once we assume  $\Lambda$  to be a scalar variable, it acquires
altogether a different status in Einstein's field equations and
its influence need not be limited only to cosmology. The solutions
of Einstein's field equations with variable  $\Lambda$  will have
a wider range and the roll of scalar  $\Lambda$ in astrophysical
problems will be of as much significance as in cosmology. \\ It is
this aspect that motivated us to reexamine the work of Ray and Ray
(1993) and Tiwari and Ray (1996) with the generalization of
anisotropic and charged source respectively. One can  realize from
the present investigations how the variable $\Lambda$ generates
different types of solutions which are physically interesting as
they provide a special class of solutions known as electromagnetic
mass models (EMMM).\\ In section 2.2,  the Einstein-Maxwell field
equations  with variable $\Lambda$ are derived. Solutions
corresponding to different cases for anisotropic system are
obtained in section 2.3. All the solutions obtained are matched
with the exterior Reissner-Nordstr\"{o}m (RN) solution on the
boundary of the charged sphere. Finally, some concluding remarks
are made in section 2.4. \\

\section{Field Equations}

The Einstein-Maxwell field equations for the spherically
symmetric metric
\bea ds^{2}= e^{\nu(r)} dt^{2} - e^{\lambda(r)}
dr^{2} - r^{2} ( d \theta ^{2} + sin^{2} \theta d\phi^{2} )
\eea
corresponding to charged anisotropic fluid distribution are given
by
\bea
  e^{-\lambda} ( \lambda^{\prime}/r - 1/r^{2} ) + 1/r^{2} = 8\pi \rho + E^{2}
+ \Lambda, \eea \bea e^{-\lambda} ( \nu^{\prime}/r + 1/r^{2} ) -
1/r^{2} = 8\pi {p}_{r} -E^{2} -\Lambda, \eea \bea
e^{-\lambda}[\nu^{{\prime}{\prime}}/2 + {\nu^{\prime}}^{2}/4 -
{\nu^{\prime} \lambda^{\prime}}/4 + (\nu^{\prime} -
\lambda^{\prime} )/ 2r] = 8\pi p_{\perp} + E^{2} - \Lambda
\eea
and
\bea {(r^2 E)}^{\prime} = 4\pi r^2 \sigma e^{\lambda/2}.
\eea
The equation (2.5) can equivalently be expressed in the form, \bea
E(r) = \frac{1}{r^2} \int_{0}^{r} 4 \pi r^2 \sigma e^{\lambda/2}
dr = \frac{q(r)}{r^2}. \eea \noindent where $q(r)$ is total charge
of the sphere under consideration.\\ Also, the conservation
equation is given by \bea \frac{d}{dr}(p_r - {\Lambda}/{8 \pi} ) +
( \rho + p_r ) {\nu^\prime}/2 = \frac{1}{8 \pi r^4}\frac{d}{dr}(
q^2) + 2( p_{\perp} - p_r)/r. \eea Here, ~$\rho$, ~$p_r$,
~$p_{\perp}$, ~$E$, ~$\sigma$ and ~$q$ are respectively the
matter-energy density, radial and tangential pressures, electric
field strength, electric charge density and electric charge. The
prime denotes derivative with respect to radial coordinate $r$
only.

 Equations (2.2) and (2.3) yield \bea
e^{-\lambda}(\nu^{\prime} + \lambda^{\prime}) = 8 \pi r( \rho +
p_r). \eea Again, equation (2.2) can be expressed in the general
form as \bea e^{-\lambda} = 1 - 2M(r)/r, \eea where \bea M(r) = 4
\pi \int_{0}^{r}[ \rho + ( E^2 + \Lambda)/{8 \pi}] r^2 dr \eea is
the active gravitational mass of a charged spherical body which is
dependent on the cosmological parameter $ \Lambda = \Lambda(r)$.\\

 \section{Solutions}

A number of solutions can be obtained depending on different
suitable conditions on equation (2.7). Here we assume the relation
$ g_{00} g_{11} = -1 $, between the metric potentials of metric
(2.1), which, by virtue of equation (2.8), is equivalent to the
equation of state \footnote[2] {In terms of energy-momentum tensor
this can be expressed as $ {T^{0}}_{0} = {T^{1}}_{1} $.} \bea \rho
+ p_r= 0. \eea The equations (2.2) -- (2.5) being underdetermined,
we further assume the following conditions \bea \sigma
e^{\lambda/2} =\sigma_{0} \eea and \bea p_{\perp}=n p_{r},\qquad
(n\neq 1), \eea where $ \sigma_{0} $ is a constant (which from
(2.6) can be interpreted as the volume density of the charge
$\sigma$ being constant) and n is the measure of anisotropy of the
fluid system.\\ Equation (2.6), with equation (2.12), provides the
electric field and charge as \bea E= q/r^{2}= 4 \pi \sigma_{0}
r/3. \eea Using equations (2.11), (2.13) and  (2.14), in equation
(2.7), we get \bea \frac {d}{dr} (p_{r}- \Lambda/{8 \pi}) - 2
(n-1) p_{r}/r= 2Ar,\quad A=2 \pi \sigma_{0}^{2}/3, \eea which is a
linear differential equation of first order. 

 Since the equation
(2.15) involves two dependent variables, $ p_{r} $ and $ \Lambda
$, to solve this equation, we consider the following four simple
cases.\\

 \subsection{$ \Lambda= \Lambda_{0}-8 \pi
p_{r},\quad (\Lambda_{0}= $constant)}

 The solutions in this case are obtained as
\bea e^{\nu}=e^{-\lambda}= 1- 2 M(r)/r, \eea \bea \rho= -p_{r}=-
p_{\perp}/n=( \Lambda-\Lambda_{0})/{ 8 \pi}=A a^{-(n-3)} r^{2}
[a^{n-3}-r^{n-3}]/(n-3) \eea and \bea M(r) =\frac {4 \pi A a
^{-(n-3)} r^{5}}{15 (n-3)(n+2)} [ (n+2)(n+3)a^{n-3}-30 r^{n-3}]+
\Lambda_{0} r^{3}/6, \eea where $ a $ is the radius of the
sphere.\\

Some general features of these solutions are as follows: \\
(1) As we want, customarily, $ \rho> 0 $ (and hence $
p_{r} < 0 $), we must have, from (2.17), $ n >3 $. However, we can
choose $ n < 3 $ (and certainly $ n \neq 1 $). In that case also
$\rho $ becomes positive. This result, viz., the positivity of
matter-energy density is obvious as the electron radius for the
present model is $~10^{-13}$~cm, which is much larger than the
experimental upper limit $10^{-16}$~cm (Quigg 1983). Within this
limit the charge distribution of matter must contain some negative
rest mass (Bonnor and Cooperstock 1989; Herrera and Varela 1994).
This is the reason why we cannot consider $ \rho \le 0 $ and hence
$ p_{r} \ge 0 $ in the present model.

(2) Similarly, we can observe that the effective gravitational mass
(which we get after matching of the interior solution to the exterior RN
solution on the boundary),
\bea
m = M(a)+ q^{2}(a)/{2 a} -
\Lambda_{0} a^{3}/6 = 8 \pi A (n+3) a^{5} /{ 5 (n+2)} ,
\eea
is positive for both the choices, $ n > 0 $ and $ n <0 $. In this
respect, the Tolman-Whittaker mass,
\bea
m_{TW}= \int_{V}
({T^{0}}_{0}-{T^{\alpha}}_{\alpha}) \sqrt{-g} dV, (\alpha= 1, 2, 3
\quad  and \quad g \rightarrow 4D)  \nonumber \\ = -\frac {8 \pi A
a ^{-(n-3)} r^{5}}{15 (n-3)(n+2)} [2(n+2)(n+3)a^{n-3}-15(n+1)
r^{n-3}] -  \Lambda_{0} r^{3}/3,
\eea
can also be examined. In general, this is negative and also equal
to modified Tolman-Whittaker mass (Devitt and Florides 1989),
\bea
m_{DF}=e^{-(\nu+\lambda)/2} m_{TW},
\eea
as $ \nu+\lambda =0 $, by virtue of the condition $ g_{00}g_{11}=-1 $
in the present paper.\\
(3) Pressure being negative the model is under
tension. This repulsive nature of pressure is associated with the
assumption (2.11), where matter-energy density is positive. This
negativity of the pressure corresponds to a repulsive
gravitational force (Ipser and Sikivie 1983; L\'{o}pez 1988).

(4) The cosmological parameter $\Lambda$, which is
assumed to vary spatially, can be shown to represent a parabola
having the equation of the form $ \Lambda = 8 \pi A [ ( a/2)^2 -
(r - a/2)^2 ] + \Lambda_{0}$ for a particular case $ n =2 $.  The
value of $ \Lambda $ increases from $ 0 $ to $a/2$ and then
decreases  from $a/2$ to $a$ and hence it is maximum at $a/2$ .
The vertex of the parabola is at $ r = a/2$ whereas the values of
$\Lambda$ at $r = 0 $ and at  $r =a$ are $\Lambda_{0}$, the
erstwhile cosmological constant. The same result can also be
obtained from equation (2.17) as at the boundary of the sphere $r =
a$, $ p_r = p_{\perp}  = 0 $ ( and hence $\Lambda = \Lambda_{0}$). 

(5) The solution set provides electromagnetic mass
model (EMMM) (Feynman et al. 1964; Tiwari et al. 1984, 1986, 1991;
Gautreau 1985; Gr{\o}n 1985, 1986a, 1986b; Ponce de Leon 1987a,
1987b, 1988; Tiwari and Ray 1991a, 1991b, 1997; Ray et al. 1993;
Ray and Ray 1993). This means that the mass of the charged
particle such as an electron originates from the electromagnetic
field alone (for a brief historical background, see Tiwari et al.
1986).

(6) The present model corresponds to Ray and
Ray (1993) for $n = 1$, under the assumption $ p_r = - \Lambda/{8
\pi}$. It can be observed that the other simple possibility, $p_r
= \Lambda/{8 \pi}$, does not exist for this case (equation (23) of
Ray and Ray (1993)). \\

\subsection{$\Lambda =\Lambda_{0} + 8 \pi p_r$}

In this case we have the following set of solutions: \bea e^\nu =
e^{-\lambda} = 1- 2M(r)/r, \eea \bea \rho = -p_r = -p_{\perp}/n =
-(\Lambda-\Lambda_{0})/{8 \pi} = Ar^2/(n-1) \eea and \bea M(r) = 4
\pi A r^5/15 + \Lambda_{0}r^3/6. \eea

Here some simple observations are as follows:\\
(1) In this
case also the electron radius being $ \sim 10^{-13} $ cm the
matter-energy density should be positive (Bonnor and Cooperstock
1989; Herrera and Varela 1994). This positivity condition requires
that $ n $ must be greater than unity.\\
(2) The effective gravitational mass,
\bea
m = 8 \pi A a^5/5,
\eea
is always positive whereas the Tolman-Whittaker mass which is also
equal to the modified Tolman-Whittaker mass, i.e.,
\bea
m_{TW} =
m_{DF} = - 16\pi A r^5/15 - \Lambda_{0} r^3/3,
\eea
is always negative in the region $ 0 <r \le a $. The gravitational mass
in this case is independent of anisotropic factor $n$.\\
(3) The pressures $ p_r $ and $ p_{\perp} $ are repulsive for $
n>1 $ (as in the previous case). \\
(4) The equation~(2.23) for $ n=2 $ can be written in the form $ \Lambda = -
8 \pi A r^2 + \Lambda_{0} $.  This yields a half-parabola whose
vertex is at $ r=0 $ and the parabola lies in the fourth-quadrant
of the coordinate systems $ ( r,\Lambda) $. \\
(5) The effective gravitational mass as obtained in (2.25) is of
 electromagnetic origin.\\
(6) The matter-energy density $ \rho $ as well as the
pressures $ p_r $ and $ p_{\perp} $ are all zero at the centre of
the spherical distribution and increase radially being maximum at
the boundary. This situation is somewhat unphysical though not at
all unavailable in the literature (Som and Bedran 1981).\\

\subsection{$ \Lambda = \Lambda_{0} - 8 \pi \int
\frac{p_r}{r} dr $}

The solution set for this case is given by \bea e^\nu =
e^{-\lambda} = 1- 2 M(r)/r, \eea \bea \rho = - p_r = - p_{\perp}/n
=2Aa^{-(2n-5)}r^2/(2n-5)[ a^{ 2n-5 } - r^{ 2n-5 }], \eea \bea
\Lambda =\frac{ 8 \pi A
a^{-(2n-5)}r^2}{(2n-3)(2n-5)}[(2n-3)a^{2n-5}-2r^{2n-5}]- 8\pi A
a^2/(2n-3)+\Lambda_{0} \eea and \bea M(r)=\frac{8\pi A
a^{-(2n-5)}r^5}{15n(2n-3)(2n-5)}[n(n+2)(2n-3)a^{2n-5}-15(n-1)r^{2n-5}]
\nonumber \\ -4\pi A a^2r^3/[3(2n-3)]+\Lambda_{0}r^3/6. \eea

Some general features of the above set of solution are
as follows: \\
(1) The matter-energy density is positive
and pressures are negative for $ n>5/2 $.\\
(2) The effective gravitational mass,
\bea
m = 8 \pi A (3n+1 ) a^{5} /15n,
\eea
is positive for $ n>1 $. On the other hand, the
Tolman-Whittaker mass and the modified Tolman-Whittaker mass,
being equal, are given by
 \bea
m_{TW} = m_{DF} =- \frac{32 \pi A a^{-(2n-5)}r^5}{15n(2n-3)(2n-5)}
[n(n+2)(2n-3)a^{2n-5}\nonumber \\ -15(n-1)(2n-1)r^{2n-5}+8 \pi A
a^{2}r^{3}/[3(2n-3)]-\Lambda_{0}r^3/3. \eea Depending on the
different values of $ n $ these masses may be negative or
positive.

 (3) The matter-energy density and the pressures, as
usual, are zero at the centre  $r=0$ as well as at the boundary $r
= a$.  Thus the maximum value must be in the region $ 0<r<a $.
This can be confirmed from equation (2.28) which, for the value $
n=2 $, represents a parabola of the form $ \Lambda = 2A
[(a/2)^2-(r-a/2)^2] $, the vertex being at $ r=a/2 $.

 (4) The
value of $\Lambda$  at the centre $ r=0 $ is $ [ \Lambda_{0} -8
\pi Aa^2/( 2n-3)]$. It acquires maximum value $ \Lambda_{0}$  at
the boundary $ r=a $.

 (5) The solution set represents EMMM.\\

\subsection{$ \Lambda = \Lambda_{0} + \int \frac
{p_r}{r} dr $}

The solutions in this case are given by
\bea
e^\nu = e^{-\lambda} = 1-2M(r)/r,
\eea
\bea
\rho = -p_r =
-p_{\perp}/n = \frac {2Aa^{-(2n-3)}r^2}{(2n-3)}[a^{2n-3}-
r^{2n-3}],
\eea
\bea
\Lambda = -\frac{ 8 \pi
Aa^{-(2n-3)}r^2}{(2n-1)(2n-3)}[(2n-1)a^{2n-3}-2r^{2n-3}]
\nonumber\\
 +8 \pi Aa^2/(2n-1) +\Lambda_{0}
\eea and \bea M = \frac{8 \pi
Aa^{-(2n-3)}r^5}{15(n+1)(2n-1)(2n-3)}[n(n+1)(2n-1)a^{2n-3}
-15(n-1)r^{2n-3}] \nonumber\\ + \frac{4 \pi Aa^2r^3}{3(2n-1)} +
\frac{\Lambda_{0}r^3}{6}. \eea

Here, the observations are as follows: \\
(1) The matter-energy density is positive and pressures are negative for
 $ n>3/2 $. \\
(2) The effective gravitational mass,
\bea
m = 8 \pi
A(3n+5)a^5/15(n+1),
\eea
for the condition $ n>1 $ is always positive, whereas the Tolman-Whittaker mass,
\bea
m_{TW} =m_{DF} =
- \frac{8 \pi
Aa^{-(2n-3)}r^5}{15(n+1)(2n-1)(2n-3)}[4n(n+1)(2n-1)a^{2n-3}
\nonumber \\-15(n-1)(2n+1)r^{2n-3}] - 8 \pi Aa^2r^3/[3(2n-1)]
-\Lambda_{0}r^3/3
\eea
may have positive or negative value depending on the choice of $ n $. \\
(3) The values related to $ \rho $ and $ p $ are zero both at $ r=0 $
and $ r=a $.\\
(4) The effective gravitational mass as well as the
other physical variables, including $ \Lambda$, are of purely
electromagnetic origin.\\

\section{Conclusions}

(1) As mentioned in the introduction, the present work
considers $ \Lambda $, the erstwhile cosmological constant, to be
a variable dependent on space coordinates. The contribution of
this variable $ \Lambda $ can be seen in the calculations given in
the previous sections. It can be seen that $ \Lambda $ is related
to pressure and matter-energy density, and therefore contributes
to effective gravitational mass of the astrophysical system. \\
(2) The present EMMMs have been obtained under the
condition $ \rho + p_r = 0 $ (equation~(2.11)). This problem thus
requires further investigation to see whether such models can be
obtained even for the condition $ \rho +  p_r \not= 0 $.\\

\vspace{5.0cm}

{\it The contents of this chapter published in}~ Indian Journal of
Pure and Applied Mathematics (2000) {\bf 31} 1017.

\pagebreak

\chapter{Classical Electron Model with Negative Energy Density
in Einstein-Cartan Theory of Gravitation}

\vspace{1.0cm}

\hrule \vspace{0.5cm}
 \begin{quote}
 {\it ``If you can look into the seeds of time, and say\\
which grain will grow and which will not..."}\\
-- Shakespear (\it Macbeth)
\end{quote}
\hrule

\vspace{1.5cm}

\section{Introduction}

Recently, Cooperstock and Rosen (1989), Bonnor and
Cooperstock (1989), and Herrera and Varela (1994) have shown that
within the experimentally obtained upper limit of the size of the
electron ( $\sim 10^{-16}$~cm) (Quigg 1983), when it is modeled as
a charged sphere obeying Einstein-Maxwell theory, must contain
some negative gravitational mass density regions within the
particle . According to Cooperstock, Rosen and Bonnor (CRB)
(1989), the rest mass or active gravitational mass within this
sphere, by virtue of the relation
\be
M = m - \frac{q^2}{2a},
\ee
is negative and about $ 10^{-52} $ cm
(when the inertial mass or effective gravitational mass, charge
and radius, respectively, of the electron, are $ m = 6.76 \times
10^{-56}$~cm, $ q = 1.38 \times 10^{-34}$~cm, and $ a = 10^{-16}
$~cm in relativistic units). Further, Herrera and Varela (HV)
(1994) have shown, in one of the cases of their paper, that the
matter-energy density
\be
\rho = (\alpha q^2 + \frac{2}{3} \pi \sigma_{0}^2)(a^2 - r^2),
\ee
for the constant $ \alpha = - 4.77 \times 10^{95} {\rm cm}^{-6} $
(when radius $ a \sim 10^{-16}$~cm) is also negative, $\sigma_{0}$
being the constant charge density at the centre of the spherical
distribution. These models,
 however, lack spin and magnetic moment and hence do not possess the
actual physical characteristics required for an electron. \\
As an
alternative way both the groups suggest the stationary Kerr-Newman
(KN) metric (Newman et al. 1965) related to the solution of
Einstein-Maxwell equations to be more appropriate than those
described earlier. However, in this context it is also to be
mentioned here that the KN metric cannot be valid for distance
scales of the radius of a subatomic particle (Mann and Morris
1993; Herrera and Varela 1994).\\
We, therefore, feel that the
problem can be tackled in the framework of Einstein-Cartan (EC)
theory, where torsion and spin are inherently present in the
formulation of the theory itself.

\section{An Overview: The Negative
Density Models}

 Before going into the Einstein-Cartan theory let us have
a bird's-eye view of the negative matter-energy density models
which we have mentioned in the introduction.\\

\subsection{The Cooperstock-Rosen-Bonnor
(CRB) Model}

Cooperstock and Rosen (1989) and Bonnor and Cooperstock
(1989) in their papers have shown that any spherically symmetric
distribution of charged fluid, irrespective of its equation of
state, whose total mass, radius and charge correspond to the
observed values of the electron, must have a negative energy
distribution (at least for some values of the radial coordinate).
Considering a static spherically symmetric charge distribution
with the line-element
\be
ds^{2}= e^{2\nu(r)}dt^{2} - e^{2 \lambda (r)} dr^{2} - r^{2}(d
\theta^{2}+sin^{2} \theta d \phi^{2})
\ee
they have argued that
when the Einstein-Maxwell equation
\be
{R^{0}}_0 - \frac{1}{2}{\delta^{0}}_0 R = 8\pi ({{T^{0}}_0}^{(m)}
+ {{T^{0}}_0}^{(em)})
\ee
is written in the form
\be
e^{-2\lambda} = e^ {2\nu} = 1 - \frac{1}{r} \int_{0}^{r}( 8\pi\rho
+ e^{-(\nu + \lambda)} E^2) r^2 dr
\ee
and hence is equated with the Reissner-Nordstr\"{o}m exterior metric
on the boundary $r = a$, which as usual gives
\be
 1 - \frac{2m}{a} + \frac{q^2}{a^2} = 1 - \frac{1}{a} \int_{0}^{a}(
8\pi\rho + e^{-(\nu + \lambda)} E^2) r^2 dr .
\ee
Then for the previous specifications of mass, charge and radius
of the electron it can be shown that
\be
\frac{q^2}{a^2} - \frac{2m}{a} \sim 2\times 10^{-36} > 0.
\ee
So, the left hand side of the above equation (3.6) must be greater than
unity and hence on the right hand side $\rho < 0 $ for some values
of $r$ implying that the electron must contain some negative rest
mass density though the net mass is as usual a positive quantity.\\

\subsection{The Herrera-Varela (HV) Model}

Following the CRB model (1989) Herrera and Varela (1994)
have discussed the fact that the electron, when modeled as a
relativistic spherically symmetric charged distribution of matter,
must contain some negative rest mass if its radius is not larger
than $\sim 10^{-16}$ cm. In this regard they have analyzed some
extended electron models and have shown that negative energy
density distributions result from the requirement that the total
mass of these models remains constant in the limit of a point
particle. Among all these extended electron models the model of
Tiwari et al. (1984) demands special attention to us which will be
seen very much relevant to our present work. Herrera and Varela
(1994) generalize this model of Tiwari et al. (1984) by
introducing a condition of anisotropy in the form
\be
p_{\perp} - p_r = \alpha q^2 r^2
\ee
 where  $\alpha$ is a constant. 

 Thus the solution obtained by Herrera and Varela (1994)
is as follows:
\be
e^{-2\lambda} = e^ {2\nu} = 1 - \frac{16}{45}{\pi}^2
{\sigma_{0}}^2 r^2 ( 5 a^2 - 2 r^2 ) - \frac{8}{15}\pi \alpha q^2
r^2 (5 a^2 - 3 r^2 ) ,
\ee
\be
p_r = - ( \alpha q^2 + \frac{2}{3}\pi {\sigma_{0}}^2) ( a^2 - r^2
) ,
\ee
\be
p_{\perp} = \alpha q^2 r^2 - (\alpha q^2 + \frac{2}{3}\pi
{\sigma_{0}}^2) ( a^2 -  r^2) ,
\ee
\be
m = \frac{64}{45} {\pi}^2 {\sigma_{0}}^2 a^5 + \frac{8}{15}\pi
\alpha q^2 a^5 ,
\ee
\be
\rho = (\alpha q^2 + \frac{2}{3}\pi {\sigma_{0}}^2) ( a^2 - r^2 )
\ee and
\be
q = \frac{4}{3}\pi \sigma_{0} a^3 \ee The value of $\alpha$ can be
obtained from the equation (3.12) as $\alpha = - 4.77 \times
10^{95}~{cm}^{-6}$ and hence the energy density, as given by the
equation (3.13) is negative for the radius of the electron $a =
10^{-16}$~cm.\\ Now, from the equation (3.12) it can be seen that
the effective gravitational mass, $m$, is of purely
electromagnetic origin and corresponds to the TRK model (1984)
with $\alpha = 0 $ case. This type of models where mass, including
all the other physical parameters, originate from the
electromagnetic field alone are known as the electromagnetic mass
models [EMMM] in the literature [Feynman et al. 1964] and have
been investigated by several  authors (Florides 1962, 1983;
Cooperstock and de la Cruz 1978; Tiwari et al. 1984, 1986, 1991,
2000; Gautreau 1985; Gr{\o}n 1985, 1986a,b; Ponce de Leon 1987a,b,
1988; Tiwari and Ray 1991a,b, 1997; Ray et al. 1993; Ray and Ray
1993). In the present paper we shall construct such a model within
the framework of Einstein-Cartan theory with negative
matter-energy density for some values of the radial coordinate.\\

\vspace{1.0cm}

\section{The Field Equations of
Einstein-Cartan Theory}

\noindent The EC field equations are given by
\be
 {R^{i}}_{j} - \frac{1}{2}{\delta^{i}}_{j} R = -\kappa {t^{i}}_{j}
\ee and
\be
{Q^{i}}_{jk} - {\delta^{i}}_{j} {Q^{l}}_{lk} -
{\delta^{i}}_{k}{Q^{l}}_{jl} = - \kappa {S^{i}}_{jk},
\ee
where $
{t^{i}}_{j} $ is the canonical energy-momentum tensor
(asymmetric), $ {Q^{i}}_{jk} $ is the torsion tensor and $
{S^{i}}_{jk} $ is the spin tensor (with $ \kappa = - 8 \pi $, ~$ G $
and ~$ c $ being unity in relativistic units).\\
The asymmetric energy-momentum tensor here is given by
\be
{t^{i}}_{j} = {T^{i}}_{j} + \frac{1}{2} g^{ik} \nabla_{m}
({S^{m}}_{jk}),
\ee
$ \nabla_{m}$  being covariant derivative with
respect to the torsionless, symmetric Levi-Civita  connection $
{\Gamma^{i}}_{jk}$ and the symmetric energy-momentum tensor ${
T^i}_j$ will consist of two parts, viz., matter and
electromagnetic tensors and which, respectively, are
\be
{{T^{i}}_{j}}^{(m)} = (\rho + p)u^{i} u_{j} - p{g^{i}}_{j} \ee and
\be
{{T^{i}}_{j}}^{(em)} = \frac{1}{4 \pi} (- F_{jk} F^{ik} +
\frac{1}{4} {\delta ^{i}}_{j} F_{kl} F^{kl} ) ,
\ee
where $ \rho $
is the matter-energy density, $p$ is the fluid pressure, $ u^{i}$
is the velocity four-vector (with $ u^{i} u_{i} = 1$) and $ F_{ij}
$ is the electromagnetic field tensor.\\
The conservation equations for the EC theory can be given through the
Bianchi identities as
\be
\nabla_{k}[(\rho + p) u^{k} - g^{ki}u^{l} \nabla_{m}(u^{m}
S_{li})] = u^{j} \nabla_{j}p \ee and
\be
[(\rho + p)u^{k} - g^{ki}u^{l} \nabla_{m}(u^{m}S^{li})]
\nabla_{k}u_{j} = - \nabla_{l}(u^{l}u_{j}) +
u^{k}S_{jm}{R^{m}}_{k} - \frac{1}{2} u^{k}S_{lm}{R^{lm}}_{jk} \ee
 Now, electromagnetic fields not being coupled with
torsion (Novello 1976; Raychaudhuri 1979) the Maxwell equations as
usual take the form
\be
\nabla _{j} F^{ij} = J^{i} \ee and

\be
( J^{i} \sqrt{- g} )_{,i} = 0. \ee The electromagnetic field
tensor, $ F_{ij}$, in the above equation (3.22) is related to the
electromagnetic potentials as $ F_{ij} = A_{i,j} - A_{j,i} $ which
is equivalent to $ F_{[i,j,k]} = 0 $, $A_{i}$ being the
electrostatic potentials. Here and in what follows a comma denotes
the partial derivative with respect to the coordinate indices
involving the index.

 Again, the spin tensor and the intrinsic
angular momentum density tensor are related in the form
\be
{S^{i}}_{jk} = u^{i}S_{jk},
\ee
with
\be
u^{i}S_{ik} = 0.
\ee
Now, assuming that the spins of the
individual charged particles composing the fluid distribution are
all aligned in the radial directions (Prasanna 1975; Raychaudhuri
1979; Tiwari and Ray 1997) and the matter is at rest with respect
to the observer, the non-vanishing components of the spin tensor
can be obtained, from equations (3.24) and (3.25), as
\be
{S^{0}}_{23} = - {S^{0}}_{32} = s(g_{00})^{- 1/2},
\ee
whereas,
from equation (3.16), we have the torsion tensor as
\be
{Q^{0}}_{23} = - {Q^{0}}_{32} = - {\kappa} s(g_{00})^{- 1/2}, \ee
$s = S_{23}$ being the only non vanishing component of the
intrinsic angular momentum density tensor. Here, we have followed
the convention $(t, r,\theta, \phi) = (0, 1, 2, 3)$.

 The
Einstein-Cartan-Maxwell equations with source can be written as
(Tiwari and Ray 1997)
\be
e^{-2 \lambda} (\frac{2 \lambda^\prime }{r} - \frac{1}{r^2}) +
\frac{1}{r^2} = 8 \pi \tilde{\rho} + E^2, \ee
\be
e^{-2 \lambda} (\frac{2 \nu^\prime}{r} + \frac{1}{r^2}) -
\frac{1}{r^2} = 8 \pi \tilde{p_r} - E^2, \ee
\be
e^{- 2\lambda} [\nu^{\prime\prime} + {\nu^{\prime}}^2
-\nu^{\prime}\lambda^{\prime}  + \frac{( \nu^{\prime}
-\lambda^{\prime})}{r}] = 8 \pi \tilde{p_\perp} + E^2 \ee and
\be
(r^2 E)^{\prime} = 4 \pi r^2 \sigma e^{\lambda},
\ee
where ~$
\tilde{\rho},~\tilde{p_r},~\tilde{p_\perp}$ and ~$E$ are the
effective matter-energy density, effective pressures (radial and
tangential) and electric field respectively, and are defined as 
\be
\tilde {\rho} = \rho -  2 \pi s^2,
\ee
\be
\tilde {p}_r = p_r - 2 \pi s^2,
\ee
\be
\tilde {p}_{\perp} = p_{ \perp} - 2 \pi s^2 \ee and
\be
E = - exp[ -(\nu + \lambda)] {\phi}^\prime = \frac{q}{r^2}, \ee
$\rho,~p_r,~p_{\perp},~s,~\phi $ and~$ q $ being the ordinary
matter-energy density, ordinary pressures (radial and tangential),
spin density, electrostatic potential and electric charge
respectively. Here, $\sigma$ represents the electric charge
density and prime denotes differentiation with respect to the
radial coordinate $r$.\\
 Now, the conservation equations
(3.20) and (3.21) with the help of equations (3.22) and (3.23)
reduce to
\be
\frac{d \tilde{p}_r}{dr} = -(\tilde {\rho} + \tilde {p}_r)
\nu^{\prime} + \frac{1}{8 \pi r^4} \frac{d}{dr} (q^2) +\frac{
2(\tilde {p}_{\perp} - \tilde {p}_r)}{r}.
\ee
This is the key
equation which is to be solved for constructing EMMM.\\

\section{The Solutions}

 Addition of (3.28) and (3.29), under the assumption $
g_{{0}{0}} g_{{1}{1}}
 = - 1 $ ( or equivalently, in terms of energy-momentum tensors
${T^0}_0 = {T^1}_1 $), provides the pure charge condition
\be
\tilde{\rho} + \tilde{p}_r = 0,
\ee
where, in general,
$\tilde{\rho} $ is assumed to be positive and hence $ \tilde {p}_r
$ is negative. However, as is evident from equation (3.32), $
\tilde{\rho}$, being the effective energy-density, can even be
negative due to the positive second term related to the spin on
the right hand side. Thus, the possibility of equation (3.33) being
satisfied with $\tilde{p_r}$ being positive is not ruled out. \\
 To make (3.31) and (3.36) solvable, we further assume that
\be
\sigma e^{\lambda}  = \sigma_{0} \ee and
\be
p_{\perp} - p_r = \tilde {p}_{\perp} - \tilde{p}_r = \alpha q^2
r^2
\ee
following TRK Model (Tiwari et al. 1984) and HV model
(Herrera and Varela 1994) respectively, where $\sigma_{0}$ and $
\alpha$ are two constants as mentioned earlier. \\
 By substituting (3.37) -- (3.39) in (3.31) and (3.36), we get
\be
E = \frac{q}{r^2}=\frac{4}{3}\pi \sigma_{0} r  ,
\ee
\be
p_r = 2 \pi s^2 - ( \alpha q^2 + \frac{2}{3}\pi {\sigma_{0}}^2) (
a^2 - r^2 ),
\ee
\be
p_{\perp} = 2 \pi s^2 - \alpha q^2 ( a^2 - 2 r^2 ) -
\frac{2}{3}\pi {\sigma_{0}}^2 ( a^2 - r^2) \ee and
\be
\rho = 2 \pi s^2 + (\alpha q^2 +\frac{2}{3}\pi {\sigma_{0}}^2) (
a^2 - r^2 ). \ee
\noindent
The active gravitational mass
\be
M(r) = 4 \pi \int_{0}^{r}( \rho - 2 \pi s^2 + \frac{E^2} {8 \pi} )
r^2 dr
\ee
takes the form, by virtue of (3.40) and (3.43), as
\be
M(r) = \frac{8}{135}{\pi}^2 {\sigma_{0}}^2 r^3 [ 8 \pi \alpha a^6
( 5 a^2 - 3r^2 ) + 3( 5a^2 - 2r^2 ) ].
\ee
 Thus, the
metric potentials $\lambda$ and $ \nu$ are given by
\be
e^{-2\lambda} = e^ {2\nu} = 1 - \frac{2M(r)}{r}, \ee whereas, the
effective gravitational mass mentioned in (3.1), can be obtained
as
\be
m =\frac{64}{45}  {\pi}^2 {\sigma_{0}}^2 a^5 (1 + \frac{2}{3}\pi
\alpha a^6 ), \ee which corresponds to the second case (B) of HV
model (1994) and is of purely electromagnetic origin. This
corresponds to the TRK model (1984) with $\alpha = 0 $ case. It
can be noted here that unlike the matter-energy density the
effective gravitational mass is independent of spin.\\ In this
context it is to be mentioned here that the junction conditions in
the EC theory are different from that of general theory of
relativity and indeed read like this (Arkuszewski et al. 1975)
\be
n_{i} u^{i} \mid _{-} = 0 \ee and
\be
p \mid _{-} = 2 \pi G ( n_{i} S^{i} ) \mid _{-} \ee where $ S^{i}$
is the spin density pseudo-vector. Here condition (3.48) is the
same as in classical relativistic hydrodynamics and has already
been incorporated by matching the interior solution with the
exterior Reissner-Nordstr\"{o}m field at the boundary of the
spherical distribution. The condition (3.49), however, is the
additional condition to be satisfied in EC theory. In the present
case, it is only the effective pressure (radial) that vanishes on
the boundary and not the ordinary radial pressure which, by virtue
of equation (3.41), equals  $ 2 \pi s^{2} $. The spin is aligned
in the radial direction and hence the spin density pseudo-vector
is hypersurface orthogonal. Thus, the boundary condition (3.49)
will become
\be
p_{r}{\mid}_{r = a - 0} = 2 \pi s^2 {\mid}_{r = a - 0},
\ee
which,
depending on whether $s$ is a constant or function of coordinates,
will automatically be satisfied.\\
 In this connection it
is to be mentioned here that the spin density, $s$, in the final
solutions (3.41) -- (3.43) remains arbitrary (function of $r$). An
explicit functional form of this spin density can also be obtained
by assuming some additional
 physically viable possibility, such as the one used by Prasanna (1975)
by splitting the conservation equation into two parts, the second
part relating to conservation of spin only, giving the functional
form of spin density as $ s = s_{0}
 e^{-\nu}$ (where $s_{0}$ is the value of $s$ at $r = 0$ i.e. the
central spin density). This can, using equations (3.38) and (3.46)
and the condition $g_{oo}g_{11} = 1$, (that is, $ \nu + \lambda =
0 $)
 equivalently be written as $ s \sigma = s_{0} \sigma_{0} =$ constant.
The functional form of spin density is, however, not relevant in
our discussion as our problem is concerned with the properties
related to the `electron', an elementary particle whose radius is
of the order of $ 10^{- 16}$~cm. Indeed,
 as the spin function is arbitrary, there is no loss in generality,
even if we assume it to be almost a constant (that is, the
quantized value of the spin of
 the electron).

\section{The Negative Energy Density
Model}

Let us have a closer observation of the results of the
previous section $3.4$. The equation (3.43) related to matter-energy
density has the spin density part in the first term where spin
density is defined as $s = 3S/4\pi a^{3}$, where $S$ is the spin
of electron the quantized value of which is $S = \hbar/2$. Then
substituting the standard values for different parameters in the
relativistic units, as mentioned in the introductory part, the
numerical value for the matter-energy density (3.43) can be shown as
\be
\rho = 6.14\times 10^{-37} - 6.81\times 10^{27} (10^{-32} - r^2).
\ee
 The first term related to spin, being of the order
of $ 10^{-37}$, is too small compared to the last term and hence
the spin contribution is negligible. Now, the equation (3.51)
indicates that the central density at $ r = 0 $ is negative and
its magnitude is about $10^{-5}$. On the other hand, the total
density at the boundary, $ r = a$, is positive as usual with the
numerical value about $10^{-37}$. This change in the sign of the
energy density is because of the presence of the spin term in
equation (3.51) which, indeed, is the contribution of the EC
theory. In the absence of spin, however, we could have negative
and zero densities at the centre and boundary of the electron
respectively.\\
 This change in the sign again indicates that the
central negative value gradually increases along the radius and
somewhere, in the region $0<r_{c}<a$, it becomes zero, where
$r_{c}$ is the critical radius. Obviously, the amount of negative
energy density is less than its positive counterpart the balance
of which ultimately provides the net density as the positive
one.\\ It is already mentioned that, in general, for any spherical
fluid distribution the density on the surface should be zero where
we are getting some non-zero value for it. This finite value of
density is solely coming from the spin contributed part $2\pi
s^{2}$. Thus for the vanishing spin the situation corresponds to
the general behaviour (vide equation (17) of Herrera and Varela
1994). In this context it is also to be noted here that up to the
critical radius behaviour of our model is similar to that of
Herrera and Varela (1994). Beyond this cut off radius the energy
density is regulated by spin which makes the overall density of
the model positive. This particular aspect lack in the model of
Herrera and Varela (1994) where the total energy density is a
negative quantity. This increase of matter-energy density due to
spin density can probably be accounted for the kinetic energy
through the angular motion of the electron here.\\ Similar kind of
examination is also possible for the pressures, radial and
tangential, both. The radial pressure, in this case of equation
(3.41), takes the following value
\be
p_r = 6.14\times 10^{-37} + 6.81\times 10^{27} ( 10^{-32} - r^2 ).
\ee However, pressure is throughout positive here from the centre
to boundary. These results, i.e. negative energy density and
positive pressure, are in accordance with the pure charge
condition (3.37) which reads as $\rho= - p_r + 4 \pi s^2$ .

\section{Conclusions}

 (i) The possible origin of the intriguing negative
matter-energy density in the work of Cooperstock and Rosen (1989),
Bonnor and Cooperstock (1989), Herrera and Varela (1994) and
present paper may be due to
 the finiteness of the total mass of the Reissner-Nordstr\"{o}m
solution (Visser 1989). Since the electrostatic energy of a point
charge is infinite, the only way to produce a finite total mass is
the presence of an infinite amount of negative energy at the
center of symmetry. According to Bonnor and Cooperstock (1989) the
negativity of the energy density and hence the active
gravitational mass is consistent with the phenomenon, known as the
Reissner-Nordstr\"{o}m repulsion (de la Cruz and Israel 1967;
Cohen and Gautreau 1979; Tiwari et al. 1984; Cooperstock and Rosen
1989). In this regards Bonnor and Cooperstock (1989) also have
discussed about the singularity theorems of general relativity
(Hawking and Ellis 1973). They have shown that the negative
regions are liable to exist over distances of order $10^{-13}$~cm
and as the proof of the singularity theorems depends on the
manifold structure of space-time valid down to lengths of order
$10^{-15}$~cm so might break down below this. On the other hand,
in the context of Einstein-Cartan theory of gravitation the idea
of negative mass is not a new one as stated by de Sabbata and
Sivaram (1994):``...torsion provides a natural framework for the
description the negative mass under extreme conditions of such as
in the early universe, when a transition from positive to negative
mass can take place.''\\ (ii) We have considered in the present
chapter an extended static spherically symmetric distribution of
an elementary particle like electron having the radius of the
order of $ 10^{- 16}$~cm, and even if for the finite size of the
physical system the spin in Einstein-Cartan theory can be related
to orbital rotation (which indeed is not the case), for systems of
dimensions of subatomic particle the orbital rotation loses its
meaning. In this case, the only way is to take the spin to be the
`intrinsic angular momentum', that is, the spin of quantum
mechanical origin (in our problem since $ s $ is arbitrary, we can
consider its quantized value or an average value). In this respect
we would like to quote here from Hehl et al. (1974),``It is
crucial to note that
 {\it{spin}} in $U_{4}$ theory is canonical spin, that is, the
{\it{intrinsic}} spin of elementary particles,
 not the so-called spin of galaxies or planets."\\
 (iii) Following other authors (Prasanna 1975;
Raychaudhuri 1979; Tiwari and Ray 1997), in the present work the
spins of all the individual particles are assumed to be oriented
along the radial axis of the spherical systems. As to how this
alignment is brought about is not very much clear. We have
discussed here only a few possible ways of realizing this
situation. According to Raychaudhuri (1979), in general, there
will be an interaction between the spins of the particles and the
magnetic field. The overall effect is the alignment of the spins.
In this context, Prasanna (1975) mentioned that such an alignment
may be meaningful either in the case of spherical symmetry when
magnetic field is present or else one has to consider axially
symmetric field. As stated above, our view point is that in the
case of physical systems of the size of the electron, the radial
alignment of the spin is not ruled out. The solution obtained
supports this view.\\
 (iv) Though our present approach
via Einstein-Cartan theory to inject spin may be interesting, we
feel even that there should have some room to discuss the
relationship of our work with an alternative means to provide spin
and magnetic moment. We think this may possible through
Dirac-Maxwell theory where spin and magnetic moment are naturally
incorporated through the Dirac spin (Bohun and Cooperstock 1999;
Lisi 1995) and would like to pursue this problem in future
investigations.\\

\vspace{2.5cm}

{\it The contents of this chapter published in} ~International
Journal of Modern Physics D (2004) {\bf 13} 555.

\pagebreak

\chapter{Energy Density in General Relativity: a Possible Role for
Cosmological Constant}

\vspace{1.0cm}

\hrule \vspace{0.5cm}
 \begin{quote}
 {\it ``The most insignificant thing contains some\\
little unknown element. We must find it!"}\\
-- Maupassant
\end{quote}
\hrule

\vspace{1.5cm}

\section{Introduction}

The structure of electron was, for a long time, an intrigue
problem to the researchers. Many scientists, like Lorentz (1904)
and even Einstein (1919) tried to solve the problem to show that
the electron mass is a electromagnetic field dependent quantity
(for a detail account see the references Tiwari, Rao and
Kanakamedala (1986) and Wilczek (1999)). Later on, under general
relativity some models have been constructed by different authors
describing extended electron with its mass entirely of
electromagnetic origin (Florides 1962; Cooperstock and de la Cruz
1978; Tiwari, Rao and Kanakamedala 1984; Gautreau 1985). Recently,
based on the experimental upper limits on the size of the electron
as $\sim 10^{-16}$~cm (1983) it is argued by Cooperstock and Rosen
(1989), Bonnor and Cooperstock (1989) and Herrera and Varela
(1994) that in the framework of general theory of relativity the
electron-like spherically symmetric charged distribution of matter
must contain some negative mass density. Being motivated by these
results with historical and heuristic values we would like to
explore a possible role for cosmological constant on the energy
density of electron when it is modeled as a dependent on the
radial coordinate $r$ of the charged spherical matter
distribution.\\ The basic logic for considering variability of so
called cosmological constant, which was introduced by Einstein in
1917 to obtain a static
 cosmological model, is related to the observational
evidence of high redshift Type Ia supernovae (Perlmutter et al.
1998; Riess et al. 1998) for a small decreasing value of
cosmological constant ($\Lambda_{present}\leq 10^{-56} cm^{-2}$)
at the present epoch. This indicates that instead of a strict
constant the $\Lambda$ could be a function of space and time
coordinates. If the role of time-dependent $\Lambda$ is prominent
in the cosmological realm, then space-dependent $\Lambda$ has an
expected effect in the astrophysical context. It is, therefore,
argued by Narlikar et al. (1991) that the space-dependence of
$\Lambda$ cannot be ignored in relation to the nature of local
massive objects like galaxies. Our aim, however, to see if there
is any effect of space-dependent $\Lambda$ on the energy density
of the classical electron. This is because cosmological constant
is thought to be related to the quantum fluctuations as evident
from the theoretical works by Zel'dovich (1967). Moreover, it is
believed through indirect evidences that 65 \% of the contents of
the universe is to be in the form of the energy of vacuum (Martins
2002). Thus, the energy density of vacuum due to quantum
fluctuation might have, in our opinion, some underlying relation
to the energy density of Lorentz's extended electron (1904) under
general relativistic treatment.\\
In the present chapter IV we have tried to find out, through some
specific case studies, that energy density of classical electron
is related to the variable cosmological constant and the
gravitational mass of the electron is entirely dependent on the
electromagnetic field alone.

\section{The field equations}

 To carry out the investigation we have considered the
 Einstein-Maxwell field equations for the case of anisotropic charged
fluid distribution (in relativistic units $G = c = 1$) which are
given by
\begin{equation}
{G^{i}}_{j} \equiv {R^{i}}_{j} - {{g^{i}}_{j}} R/2 = -8\pi
[{{T^{i}}_{j}}^{(m)}+{{T^{i}}_{j}}^{(em)}+ {{T^{i}}_{j}}^{(vac)}
],
\end{equation}
\begin{equation}
{[{(-g)}^{1/2}F^{ij}], }_{j}= 4\pi J^{i}{(-g)}^{1/2}
\end{equation}
and
\begin{equation}
F_{[ij, k]}= 0
\end{equation}
where ${F^{ij}}$ is the electromagnetic field tensor and
${J^{i}}$,
 current four vector which is equivalent to ${J^{i}}= \sigma {u^{i}}$,
 $\sigma $ being the charge density and $u^{i}$ is the four-velocity
 of the matter satisfying the relation $u_{i}u^{i} = 1$.\\
The matter, electromagnetic and vacuum energy-momentum tensors
are, respectively given by
\begin{equation}
{{T^{i}}_{j}}^{(m)} = (\rho + p_{\perp}) u^{i}u_{j} - p_{\perp} {
g^{i}}_{j} +(p_{\perp} -p_{r}){\eta}^{i}{\eta}_{j},
\end{equation}
\begin{equation}
{{T^{i}}_{j}}^{(em)}= - [ F_{jk}F^{ik} -
{g^{i}}_{j}F_{kl}F^{kl}/4]/4\pi
\end{equation}
and
\begin{equation}
 {{T^{i}}_{j}}^{(vac)}= {{g^i}_j}\Lambda(r)/8\pi
\end{equation}
where  $\rho$,  $p_r$ and  $p_{\perp}$  are the proper energy
density, radial and tangential pressures respectively and $\eta_i$
is the unit space-like vector on which the condition to be imposed
is ${\eta}_{i}{\eta}^{i} = - 1$. Here $p_r$ is the pressure in the
direction of $\eta_i$ whereas $p_{\perp}$ is the pressure on the
two-space orthogonal to $\eta_i$.\\ Now, for the spherically
symmetric metric
\begin{equation}
ds^2 = e^{\nu(r)}dt^2 - e^{\lambda(r)}dr^2 - r^2(d{\theta^2} +
 sin^2{\theta}d{\phi}^2),
\end{equation}
the Einstein-Maxwell field equations (4.1) - (4.6) corresponding
to anisotropic charged fluid with spatially varying cosmological
constant i.e. $\Lambda = \Lambda(r)$, are given by
\begin{equation}
e^{-\lambda}(\lambda^\prime/r - 1/r^2) + 1/r^2 = 8\pi T^0_0 =
8\pi{\tilde {\rho}} + E^2,
\end{equation}
\begin{equation}
e^{-\lambda}({\nu}^{\prime}/r + 1/r^2) - 1/r^2 = - 8\pi{{T^1}_1} =
8\pi \tilde{ p_r} - E^2,
\end{equation}
$$\hspace{-2in}e^{-\lambda}[{\nu}^{{\prime}{\prime}}/2 +
{{\nu}^{\prime}}^2/4 - {{\nu}^{\prime}}{{\lambda}^{\prime}}/4 +
({\nu}^{\prime} - {\lambda}^{\prime})/2r]$$
\begin{equation}
~~~~~~~~~~~~~~~~~~= - 8\pi{{T^2}_2} = - 8\pi{{T^3}_3} = 8\pi
\tilde {p_{\perp}} + E^2
\end{equation}
and
\begin{equation}
[r^2 E]^{\prime} = 4\pi r^2 {\sigma} e^{{\lambda}/2},
\end{equation}
where $E$, the intensity of electric field, is defined as $ E =
-e^{-(\nu + \lambda)/2}{\phi}^{\prime}$ and can equivalently be
expressed, from equation (4.11), as
\begin{equation}
E = \frac{1}{r^2}\int^{r}_{0} 4\pi r^2 {\sigma} e^{\lambda/2} dr.
\end{equation}
Here prime denotes differentiations with respect to the radial
coordinate $r$ only.\\
 In the above equations (4.8) -- (4.10) we have considered that
\begin{equation}
\tilde {\rho}  = \rho + \Lambda(r)/8\pi,
\end{equation}
\begin{equation}
{\tilde p}_r = p_r - \Lambda(r)/8\pi
\end{equation}
and
\begin{equation}
{\tilde p}_{\perp} = p_{\perp} - \Lambda(r)/8\pi,
\end{equation}
where ${\tilde {\rho}}$, ${{\tilde p}_r}$ and ${{\tilde
p}_{\perp}}$ are the effective energy density, radial and
tangential pressures respectively.\\
 The equation of continuity ${{T^i}_j};i = 0$, is given by
\begin{equation}
\frac{d{p_r}}{dr} - \frac{1}{8\pi}\frac{d\Lambda(r)}{dr} +
\frac{1}{2}(\rho+p_r){\nu^\prime} =\frac{1}{8\pi
r^4}\frac{dq^2}{dr} + \frac{2(p_{\perp}-p_r)}{r}
\end{equation}
where $q$ is the charge on the spherical system.\\
We assume the relation between the radial and tangential pressures
(Herrera and Varela 1994) as
\begin{equation}
 p_{\perp} - p_r = \alpha q^2r^2,
\end{equation}
where $\alpha$ is a constant.\\
Hence, by use of equations (4.14) and (4.17), the equation (4.16)
reduces to
\begin{equation}
\frac{d \tilde {p_r}}{dr}+\frac{1}{2}( \tilde {\rho} +
 {{\tilde p}_r}){{\nu}^{\prime}}
 =\frac{1}{8\pi r^4}\frac{dq^2}{dr}+2{\alpha}q^{2}r.
\end{equation}
Now, equation (4.8) can be expressed in the following form as
\begin{equation}
 e^{-\lambda}= 1 - 2M/r,
\end{equation}
where the active gravitational mass, $M$, is given by
\begin{equation}
M = 4{\pi}{\int_0^r}\left[\tilde {\rho}
+\frac{E^2}{8{\pi}}\right]r^2dr.
\end{equation}

\section{The solutions}

\subsection{Model for $\rho +p_r = 0$}

Let us now solve the equation (4.18) under the assumption
between the stress-energy tensors as ${T^1}_1 = {T^0}_0$, which
implies that
\begin{equation}
\tilde{\rho} + \tilde{ p_r} = \rho + p_r = 0.
\end{equation}
In order to make the equation (4.12) integrable we assume that
\begin{equation}
\sigma = {\sigma}_0 e^{-\lambda/2},
\end{equation}
where $\sigma_0$ is the charge density at $r = 0$ of the spherical
distribution, i.e. the central density of charge.

 Now, using
condition (4.22) in equation (4.12), we get for the expression of
electric charge and intensity of the electric field as
\begin{equation}
q = Er^2 = \frac{4}{3}\pi{\sigma}_0r^3.
\end{equation}
With the help of equations (4.21) and (4.23), the equation (4.18)
reduces to
\begin{equation}
\frac{d \tilde {p_r}}{dr}=\frac{4}{3}\pi{\sigma_0}^2r+2{\alpha}
q^2r,
\end{equation}
Thus the  solution set is given by
\begin{equation}
e^{-\lambda}=e^{\nu}=1-\frac{16}{45}{\pi}^2{\sigma_0}^2r^2(5a^2-2r^2)-\frac{8}{15}\pi\alpha
q^2r^2(5a^2-3r^2),
\end{equation}
\begin{equation}
p_{r}= - (\alpha q^2 +\frac{2}{3}\pi\sigma^2_0)(a^{2}-r^{2}) +
\frac{\Lambda(r)}{8\pi},
\end{equation}
\begin{equation}
p_{\perp}= \alpha q^2r^2 - (\alpha q^2 +
\frac{2}{3}\pi\sigma^2_0)(a^2-r^2)+\frac{\Lambda(r)}{8\pi}
\end{equation}
and
\begin{equation}
\rho(r)=(\alpha q^2 + \frac{2}{3}\pi\sigma^2_0)(a^2-r^2)-
\frac{\Lambda(r)}{8\pi}.
\end{equation}
The active gravitational mass which is defined in the equation
(4.20), then, by virtue of the equations (4.23) and (4.28), takes the
form as
\begin{equation}
M(r) = \frac{8}{135}{\pi}^2 {\sigma_{0}}^2 r^3 [8 \pi \alpha a^6 (
5 a^2 - 3r^2 ) + 3( 5a^2 - 2r^2 )].
\end{equation}
Thus, the metric potentials $\lambda$ and $ \nu$ are given by
\begin{equation}
e^{-\lambda} = e^ {\nu} = 1 - \frac{2M(r)}{r}.
\end{equation}
The total effective gravitational mass can be obtained, after
smoothly matching of the interior solution to the exterior
Reissner-Nordstr{\"o}m solution on the boundary, as
\begin{equation}
m = M(a) + \frac{q^2(a)}{2a} = \frac{64}{45} {\pi}^2
{\sigma_{0}}^2 a^5 (1 + \frac{2}{3}\pi \alpha a^6),
\end{equation}
which corresponds to the second case (B) of Herrera-Varela model
(1994) and represents ``electromagnetic mass'' model such that
gravitational mass of a charged fluid sphere originates from the
electromagnetic field alone (Lorentz 1904; Feynman et al. 1964).
This again corresponds to the Tiwari-Rao-Kanakamedala model (1984)
with $\alpha = 0 $ case and thus the present model reduces to
isotropic one.\\
Now, considering the observed values of mass, charge and radius of
the electron (in relativistic units) as $m=6.76\times10^{-56}$ cm,
$q=1.38\times 10^{-34}$ cm and $a=10^{-16}$ cm the value of
$\alpha$, from the equation (4.31), is given by
\begin{equation}
\alpha=-4.77\times10^{95}cm^{-6}.
\end{equation}
For the above value of $\alpha$, the energy density in equation
(4.28) becomes
\begin{equation}
\rho(r)= - 6.81 \times 10^{27}(a^2 - r^2) -
\frac{\Lambda(r)}{8\pi}.
\end{equation}
The central energy density, ${\rho}_0$, at $r=0$, then can be
calculated as
\begin{equation}
{\rho}_0 = - 6.81\times10^{-5} - \frac{\Lambda_0}{8\pi}.
\end{equation}
Thus, from the equation~(4.34) one can see that for $\Lambda_0> 0$
the energy density of the electron is a negative quantity. It is
to be noted here that in the cosmological context $\Lambda$
positive is related to the repulsive pressure and hence an
acceleration dominated universe as suggested by the SCP and HZT
project report (Perlmutter et al 1998; Riess et al. 1998;
Filippenko 2001; Kastor and Traschen 2002). However,
equation~(4.34) indicates that this negativity of energy density
is also obtainable for $\Lambda_0 < 0$ (which indicates a
collapsing situation of the universe (Cardenas et al. 2002)) for
its very small value. In this context it is also possible to show
that at an early epoch of the universe when the numerical value of
negative $\Lambda$ is higher than that of the first term of $\rho$
(i.e. $\sim 10^{-5}$ at $r = 0$) obviously energy density is a
positive quantity. Thus, in the case of decreasing negative value
of $\Lambda$ it is clear that there is a smooth crossover from
positive energy density to a negative energy density via a phase
of null energy density! However, these results confirm the vacuum
equation of state $\rho + p_r = 0$ (Davies 1984; Blome and
Priester 1984; Hogan 1984; Kaiser and Stebbins 1984).

 We can
also see that on the boundary, $r = a$, the total energy density
becomes
\begin{equation}
\rho_a = -\frac{\Lambda_a}{8\pi},
\end{equation}
which shows its clear dependency on the cosmological constant.
However, for $\Lambda_a > 0$, $\rho_a$ is  negative whereas for
$\Lambda_a < 0$, $\rho_a$  is as usual a positive quantity.
Through a simple and interesting exercise (as all the parameters
related to the electron are known) one can find out the numerical
value of $\Lambda_a$, at the boundary of the spherical system from
the equation (4.35), which equals $\sim 10^{-7} cm^{-2}$. This
constant value of $\Lambda_a$ is too large and might be related to
an early epoch of the universe. Here for finding out the total
energy density $\rho_a$ it is considered that $\rho_a \leq
\rho_{average}$, where $\rho_{average}$ is equal to
$m/\frac{4}{3}\pi a^3$ as the energy density of the spherical
distribution is decreasing from centre to boundary.\\

\subsection{Model for $\rho + p_r \neq 0$}

Now using equation (4.23) the equation (4.18) can be
written as
\begin{equation}
\frac{d}{dr}\left[\tilde{p_r} - \frac{E^2}{8\pi}\right]+
\frac{1}{2}(\tilde{\rho}+ \tilde{p_r}){{\nu}^{\prime}} = \frac{
E^2}{2\pi r} + 2\alpha q^2r.
\end{equation}
Assuming that the radial stress-energy tensor ${T^1}_{1}=0$, one
gets
\begin{equation}
{\nu}^{\prime}= \frac{(e^{\lambda}-1)}{r},
\end{equation}
and
\begin{equation}
{\tilde p}_r= \frac{E^2}{8\pi}.
\end{equation}
Using equations (4.37) and (4.38) in equation (4.36), we have
\begin{equation}
\tilde{\rho}+ {\tilde p}_r= \rho + p_r =\frac{(4\alpha q^2 r^2+
E^2/\pi)}{e^\lambda -1}.
\end{equation}
\noindent Thus, equation (4.20) takes the form as
\begin{equation}
M = 4\pi\int_{0}^{r}\left[\frac{(4\alpha q^2 r^2
+E^2/\pi)}{e^\lambda -1}\right]r^2dr.
\end{equation}
 To make equation (4.40) integrable we assume that
\begin{equation}
E^2 = \pi k (e^\lambda -1)(1-R^2)-4\pi\alpha q^2 r^2,
\end{equation}
where $k$ is a constant and $R=r/a$, $a$ being the radius of the
sphere.\\
Thus, the solution set is given by
\begin{equation}
e^{-\lambda} = 1- A R^2 (5 - 3R^2),
\end{equation}
\begin{equation}
e^\nu = (1-2A)^{5/4}e^{\lambda/4}exp[5B{tan^{-1}B(6R^2-5)-
\frac{1}{2}tan^{-1}B}],
\end{equation}
\begin{equation}
p_r = \frac{1}{8}k(e^\lambda-1)(1-R^2) - \frac{1}{2}\alpha q^2 r^2
+ \frac{\Lambda(r)}{8\pi},
\end{equation}
\begin{equation}
p_{\perp} = \frac{1}{8}k(e^\lambda-1)(1-R^2) + \frac{1}{2}\alpha
q^2 r^2 + \frac{\Lambda(r)}{8\pi},
\end{equation}
and
\begin{equation}
\rho = k(1-R^2)[1 - \frac{1}{8}(e^\lambda -1)] + \frac{1}{2}\alpha
q^2 r^2
 - \frac{\Lambda(r)}{8\pi},
\end{equation}
where the constant $A = 8\pi ka^2/15$.\\
By application of the matching condition at the boundary we again
get the total effective gravitational mass, which in the present
case takes the form
\begin{equation}
m = \frac{8}{15}\pi k a^3 + \frac{q^2}{2a}.
\end{equation}
In view of the equation (4.41), for vanishing charge the constant
$k$ vanishes and hence makes the gravitational mass in the
equation (4.47) to vanish. Thus, the present case $\rho + p_r = k(1
- R^2) \neq 0$ also represents electromagnetic mass model. \\
Now,
the constant $k$ can be expressed in terms of the known values of
the electric mass, radius and charge as
\begin{equation}
k = \frac{15}{16\pi a^4}(2am - q^2).
\end{equation}
At $r=0$, the energy density, from the equation (4.46), is then
given by
\begin{equation}
\rho_0 = - 5.68 \times 10^{-5} - \frac {\Lambda_0}{8\pi}.
\end{equation}
As, in the case of electron, $k$ is a negative quantity so for
$\Lambda_0 > 0$ the central energy density $\rho_0$ is negative
only. However, for $\Lambda_0 < 0$ the central energy density may
respectively be negative and positive depending on the numerical
value of $k$ whether it is higher and lower than that of
$\Lambda_0$.\\
At $r = a$, the total energy density is given by
\begin{equation}
\rho_a = - 4.54 \times 10^{-5} - \frac{\Lambda_a}{8\pi}.
\end{equation}
Similarly, for $\Lambda_a>0$, the energy density is negative
whereas for $\Lambda_a<0$, it may either be negative or positive
depending on the numerical value of $\Lambda_a$ as discussed in
the previous case.

\subsection{A test model}

In the previous two cases we have qualitatively
discussed the effect of cosmological parameter $\Lambda(r)$ on the
energy density $\rho(r)$ of the electron. Let us now explore some
quantitative effect and hence treat the equation (4.24) in  a
different way. If we substitute the value of $\tilde p_r$, from
equation (4.14), then integrating equation (4.24) we get
\begin{equation}
\Lambda_{eff} = \Lambda(a) - \Lambda(r) = 8\pi \rho(r) -
8\pi(\alpha {q^2} + \frac{2}{3}\pi\sigma^2_0)(a^2 - r^2),
\end{equation}
where $\Lambda_{eff}$ is the effective cosmological parameter.\\
We study the following cases:\\
For the central value of the
energy density of the spherical distribution, i.e. at $r=0$, the
effective cosmological parameter becomes
\begin{equation}
\Lambda_{eff}^0 = \Lambda(a) - \Lambda(0) = 8\pi \rho_0 -
8\pi(\alpha {q^2} + \frac{2}{3}\pi\sigma^2_0)a^2.
\end{equation}
Considering that $\rho_0 \geq \rho_{average}$ the effective
cosmological parameter, at $r = 0$, for the proper numerical
values of the charge and radius of the electron can be found out
as
\begin{equation}
\Lambda_{eff}^0 =  1.71 \times 10^{-3} cm^{-2}.
\end{equation}
On the other hand, at the boundary, $r=a$, of the spherical
distribution the effective cosmological parameter becomes
\begin{equation}
\Lambda_{eff}^a =\Lambda(a) - \Lambda(a) = 0.
\end{equation}
Thus, from the equations (4.53) and (4.54) it is shown that the
effective cosmological parameter has a finite value at the centre
of the electron which decreases radially and becomes zero at the
boundary.

\section{Discussions}

 We see from the above analysis that the cosmological
parameter $\Lambda$ has a definite role even on the energy density
of micro-particle, like electron. We, therefore, feel that it may
also be possible to extrapolate the present investigation to the
massive astrophysical bodies to see the effect of spatially
varying cosmological parameter on their energy densities and vice
versa.\\
The proper pressure $p_r$, in general, being positive as
evident from the equation (4.26) is in accordance with the
condition (4.21) which may be explained as due to vacuum
polarization (Gr{\o}n 1985). In this connection it is mentioned by
Bonnor and Cooperstock (1989) that the negativity of the active
gravitational mass and hence negative energy density for electron
of radius $a \sim 10^{-16}$ is consistent with the
Reissner-Nordstr{\"o}m repulsion. We would also like to mention
here that the equation of state in the form $p + \rho = 0$ is
discussed by Gliner (1966) in his study of the algebraic
 properties of the energy-momentum tensor of ordinary matter through the
metric tensors and called it the $\rho$-vacuum state of matter. It
is also to be noted that the gravitational effect of the
zero-point energies of particles and electromagnetic fields are
real and measurable, as in the Casimir Effect (1948). According to
Peebles and Ratra (2002), like all energy, this zero-point energy
has to contribute to the source term in Einstein's gravitational
field equation. This, therefore, demands inclusion of vacuum
energy related term cosmological constant in the field equation.
In this regard it is interesting to recall the comment made by
Einstein (1919) where he stated that ``... of the energy
constituting matter three-quarters is to be ascribed to the
electromagnetic field, and one-quarter to the gravitational
field'' and did ``disregard'' the cosmological constant in his
field equation is in contradiction to the present result as shown
in the equation (4.31) and (4.47).

\vspace{5.0cm}

{\it The contents of this chapter published in} ~Physics Letters A
(2004) {\bf 322} 150.

\pagebreak

\chapter{Relativistic Electromagnetic Mass Models: Charged Dust
Distribution in Higher Dimensions}

\vspace{1.0cm}

\hrule \vspace{0.5cm}
 \begin{quote}
 {\it ``The most incomprehensible thing about\\
 the world is that it is comprehensible."}\\
-- Albert Einstein
\end{quote}
\hrule

\vspace{1.5cm}

\section{Introduction}

In many theories higher dimensions play an important
role, specially in super string theory (Schwarz 1985; Weinberg
1986) which demands more than usual four dimensional space-time.
This is also true in studying the models regarding unification of
gravitational force with other fundamental forces in nature. In
the case of a simple solution to the vacuum field equations of
general relativity in $(4+1)$ space-time dimensions Chodos and
Detweiler (1980) have shown that it leads to a cosmology which at
the present epoch has $(3+1)$ observable dimensions in which the
Einstein-Maxwell equations are obeyed. Lorenz-Petzold (1984) has
studied a class of higher dimensional Bianchi-Kantowski-Sachs
space-times of the Kaluza-Klein type whereas Iban{\`e}z and
Verdaguer (1986) have examined radiative isotropic cosmologies
with extra dimensions related to FRW models.\\
In this connection it is interesting to note that electromagnetic
mass models where all the physical parameters, including the
gravitational mass, are arising from the electromagnetic field
alone have been extensively studied (Tiwari et al. 1984; Gautreau
1985; Gr{\o}n 1986; Ponce de Leon 1987; Tiwari and Ray 1991a; Ray
and Bhadra 2004a,b) in the space-time of four dimensional general
relativity. Thus it is believed that study of electromagnetic mass
models in higher dimensional theory will be physically more
interesting.\\
Under this motivation we have considered here a static spherically
symmetric charged dust distribution corresponding to higher
dimensional theory of general relativity. It is proved, as a
particular case, from the coupled Einstein-Maxwell field equations
that a bounded and regular interior static spherically symmetric
charged dust solution, if exists, can only be of purely
electromagnetic origin. An example, which is already available, is
examined in this context and is shown that the solution set
satisfies the condition of being electromagnetic origin.

\section{The Einstein-Maxwell Field Equations}

 The Einstein-Maxwell field equations for the case of
charged dust distribution are given by
\begin{eqnarray}
{G^{i}}_{j} \equiv {R^{i}}_{j} - \frac{1}{2}{{g^{i}}_{j}} R =
-8\pi [{{T^{i}}_{j}}^{(m)}+{{T^{i}}_{j}}^{(em)}],
\end{eqnarray}
\begin{eqnarray}
{[{(-g)}^{1/2}F^{ij}], }_{j}= 4\pi J^{i}{(-g)}^{1/2}
\end{eqnarray}
and
\begin{eqnarray}
F_{[ij, k]}= 0
\end{eqnarray}
 where ${F^{ij}}$ is the electromagnetic field tensor and
${J^{i}}$ the current four vector which is equivalent to
\begin{eqnarray}
{J^{i}}= \sigma {u^{i}}
\end{eqnarray}
$\sigma $ being the charge density and $u^{i}$ is the four
velocity of the matter satisfying the relation
\begin{eqnarray}
u_{i}u^{i}= 1.
\end{eqnarray}
The matter and electromagnetic energy momentum tensors are given
by
\begin{eqnarray}
{{T^{i}}_{j}}^{(m)}= \rho u^{i}u_{j}
\end{eqnarray}
and
\begin{eqnarray}
{{T^{i}}_{j}}^{(em)}= \frac{1}{4\pi}[- F_{jk}F^{ik}
+\frac{1}{4}{g^{i}}_{j}F_{kl}F^{kl}],
\end{eqnarray}
where $\rho $ is the proper energy density.\\
  Now we consider the $(n+2)$ dimensional spherically symmetric metric
\begin{eqnarray}
ds^{2}= e^{\nu(r)}dt^{2}- e^{\lambda(r)}dr^{2}- r^{2}{X_{n}}^{2}
\end{eqnarray}
where
\begin{eqnarray}
{X_{n}}^{2}=
d{\theta_{1}}^{2}+sin^{2}{\theta_{1}}d{\theta_{2}}^{2}+\nonumber
sin^{2}{\theta_{1}}sin^{2}{\theta_{2}}d{\theta_{3}}^{2}+......\nonumber
+\left[\prod^{n-1}_{i=1}sin^{2}{\theta_{i}}\right]d{\theta_{n}}^{2}.
\end{eqnarray}
The convention adopted here for coordinates are
\begin{eqnarray}
x^{1}= r,\quad x^{2}=\theta_{1},\quad
x^{3}=\theta_{2},.....x^{n+1}=\theta_{n},\quad x^{n+2}= t
\end{eqnarray}
and also
\begin{eqnarray}
g_{11}=-e^\lambda,\quad g_{22}= -r^{2},\quad
g_{33}=-r^{2}sin^{2}{\theta_{1}},\nonumber\\
g_{44}=-r^{2}sin^{2}{\theta_{1}}sin^{2}{\theta_{2}},.....
g_{(n+1)(n+1)}=-r^{2}\left[\prod^{n-1}_{i=1}{sin^{2}{\theta_{i}}}\right],\nonumber\\
g_{(n+2)(n+2)}= e^\nu.
\end{eqnarray}
As we have considered here a static fluid, so
\begin{eqnarray}
u^{i}= [0,0,0,.....(n+1)times,\quad e^{-{\nu}/{2}}]
\end{eqnarray}
and
\begin{eqnarray}
J^{1}= J^{2}= J^{3}=...J^{n+1}= 0,\quad J^{n+2}\neq 0
\end{eqnarray}
so that the only non-vanishing components of $F_{ij}$ of equations
${5.2}$ and ${5.3}$ are $F_{{1}(n+2)}$ and $F_{(n+2){1}}$.

 In
view of the above, the Einstein-Maxwell field equations for static
spherically symmetric charged dust corresponding to the metric
(5.8) are
\begin{eqnarray}
e^{-\lambda}[n{\nu^{\prime}/{2r}}+n(n-1)/2r^{2}]-n(n-1)/2r^{2}=
-E^{2},
\end{eqnarray}
\begin{eqnarray}
e^{-\lambda}[\nu^{{\prime}{\prime}}/2+{\nu^{\prime}}^{2}/4-
{\nu^{\prime}\lambda^{\prime}}/4+(n-1)(\nu^{\prime}-\lambda^{\prime})/2r+\nonumber
\\ (n-1)(n-2)/2r^{2}] -(n-1)(n-2)/2r^{2}= E^{2},
\end{eqnarray}
\begin{eqnarray}
e^{-\lambda}[n{\lambda^{\prime}}/2r-n(n-1)/2r^{2}]+n(n-1)/2r^{2}=
8\pi\rho+E^{2}
\end{eqnarray}
and
\begin{eqnarray}
[r^{n}E]^{\prime}= 4\pi r^{n} \sigma e^{\lambda/2}
\end{eqnarray}
where $E$, the electric field strength, is defined as $ E=
-e^{-(\nu+\lambda)/2}\phi^\prime$ , the electrostatic potential
$\phi$  being related to the electromagnetic field tensor as
$F_{(n+2)1}= -F_{1(n+2)}=\phi^{\prime}$.

\section{Higher Dimensional
Electromagnetic Mass Models}

>From the field equations (5.13) and (5.15), we have
\begin{eqnarray}
e^{-\lambda}(\nu^\prime+\lambda^\prime)= 16\pi r\rho/n.
\end{eqnarray}
Now, from the conservation equations ${{T^{i}}_{j}};i= 0$ one
obtains
\begin{eqnarray}
\rho{\nu^\prime}= [q^{2}]^\prime/4\pi r^{4} + (n-2)E^2/2\pi r
\end{eqnarray}
where the charge, $q$, is related with the electric field
strength, $E$, through the integral form of the Maxwell's equation
(5.16), which can be written as
\begin{eqnarray}
q = Er^{n} = 4\pi\int_{0}^{r}\sigma e^{\lambda/2}r^{n}dr.
\end{eqnarray}
Again,  equation (5.15) can be expressed in the following form as
\begin{eqnarray}
e^{-\lambda}= 1- 4M/nr^{n-1}
\end{eqnarray}
where the active gravitational mass, $M$, is given by
\begin{eqnarray}
M = 4\pi\int_{0}^{r}[\rho+E^{2}/8\pi]r^{n}dr.
\end{eqnarray}
Hence, following the technique of Tiwari and Ray (1991a) we see
that for the vanishing charge density, $\sigma$, of the equation
(5.19) one can arrive, through the equation (5.18), at the unique
relation
\begin{eqnarray}
\rho\nu^\prime= 0.
\end{eqnarray}

Thus, we have the following two cases:\\
{\it Case I:
$\rho \ne0$, \quad $\nu^\prime= 0$}\\
 For this case,
from equations (5.13) and (5.14), we have $\lambda$ as a constant
quantity. This in turn makes $\rho$ equal to zero and hence by
virtue of equation (5.22) space-time becomes flat.\\

{\it Case II: $\rho=0$,\quad $\nu^\prime\ne0$}\\
 In
this case, from equations (5.20) and (5.21), $\lambda$ becomes
zero. Then from equation (5.17), we have the metric potential as a
constant and again the space-time becomes flat.\\ We are not
considering here the third case, viz. $\rho= \nu^\prime= 0$, which
is quite a trivial one. However, from the above cases (I) and (II)
it is evident that, at least at a particular case, all the charged
dust models are of electromagnetic origin, viz., all the physical
parameters are originating purely from electromagnetic field. This
type of models are known as electromagnetic mass models in the
literature (Lorentz 1904; Feynman et al. 1964).

\vspace{1.0cm}
{\bf  An example}:\\

 The solution set obtained by Khadekar et al. (2001) for
the static spherically symmetric charged dust is as follows:
\begin{eqnarray}
e^\nu = Ar^{2N},
\end{eqnarray}
\begin{eqnarray}
e^{-\lambda} =\left[\frac{(n-1)}{N+(n-1)}\right]^{2},
\end{eqnarray}
\begin{eqnarray}
\rho =\frac{Nn(n-1)^{2}}{8 \pi r^{2}[N+(n-1)]^{2}}
\end{eqnarray}
and
\begin{eqnarray}
\sigma = \frac{N(n-1)^{2}[n(n-1)]^{1/2}}{4 \pi
2^{1/2}r^2[N+(n-1)]^2}.
\end{eqnarray}
where $A$ and $N$ both are constants with the restriction that
$N\geq0$.\\
The total charge and mass of the sphere in terms of
its radius, $a$, are respectively given by
\begin{eqnarray}
q = \frac{N[n(n-1)]^{1/2}a^{n-1}}{2^{1/2}[N+(n-1)]}
\end{eqnarray}
and
\begin{eqnarray}
m =\frac{Na^{n-1}}{N+(n-1)}.
\end{eqnarray}
The charge and mass densities in the present case take the
relationship
\begin{eqnarray}
\sigma = [2(1-1/n)]^{1/2}\rho.
\end{eqnarray}
Therefore, the charge and mass densities are proportional to each
other with the constant of proportionality $[2(1-1/n)]^{1/2}$
which takes the value unity for $n=2$ i.e. in the four dimensional
case and the relation (5.29) reduces to the usual form
\begin{eqnarray}
\sigma = \pm\rho
\end{eqnarray}
which is known as the De-Raychaudhuri (1968) condition for
equilibrium of a charged dust fluid.\\
Thus, from equations
(5.23) -- (5.29), it is evident that all the physical quantities
including the effective gravitational mass vanish and also the
spherically symmetric space-time becomes flat when the charge
density vanishes implying $N=0$. The solution here, therefore,
satisfies the criteria of being of purely electromagnetic
origin.

\section{Discussions}

The present paper is, in general, higher dimensional analogue of
the work of Tiwari and Ray (1991a) whereas the example given here
(Khadekar et al. 2001) is the higher dimensional analogue of the
paper of Pant and Sah (1979). Thus, we have presented here a model
which corresponds to spherically symmetric gravitational sources
of purely electromagnetic origin in the space-time of higher
dimensional theory of general relativity. It has been already
proved in the four dimensional case of the present paper (Tiwari
and Ray 1991a) that a bounded continuous static spherically
symmetric charged dust solution, if exists, can only be of
electromagnetic origin. Hence, this is also true in the higher
dimensional case in general theory of relativity.\\ In this regard
we would like to discuss briefly the role of higher dimensions in
different context. It have been shown by Iban{\`e}z and Verdaguer
(1986) that for the open models related to FRW cosmologies the
extra dimensions contract as a result of cosmological evolution
whereas for flat and closed models they contract only when there
is one extra dimension. Fukui (1987) recovers Chodos-Detweiler
(1980) type solutions, as mentioned in the introduction, where the
Universe expands as $t^{1/2}$ by the percolation of radiation into
4D space-time from the fifth dimension, mass, although the 5D
space-time-mass Universe itself is in vacuum as a whole. It is
interesting to note that considering mass as fifth dimension a lot
of other works also have been done by several researchers (Wesson
1983; Banerjee, Bhui and Chatterjee 1990; Chatterjee and Bhui
1990) which contain Einstein's theory embedded within it. In one
of such investigations it is argued that a huge amount of entropy
can be produced following shrinkage of extra-dimension which may
account for the very large value of entropy per baryon observed in
4D world (Chatterjee and Bhui 1990). Kaluza-Klein type higher
dimensional inflationary scenario have been discussed by Ishihara
(1984) and Gegenberg and Das (1985) where it is shown that the
contraction of the internal space causes the inflation of the
usual space.

\vspace{13.0cm}

{\it The contents of this chapter published in} ~ Astrophysics and
Space Science (2006) {\bf 302} 153.

 \pagebreak

\chapter{Relativistic Anisotropic Charged Fluid Spheres with
Varying Cosmological Constant}

\vspace{1.0cm}

\hrule \vspace{0.5cm}
 \begin{quote}
 {\it ``What are man's truths ultimately? \\
Merely his irrefutable errors!"}\\
-- Nietzsche
\end{quote}
\hrule

\vspace{1.5cm}
\section{Introduction}

The cosmological constant $\Lambda$, related to the energy of
space, introduced by Einstein in general relativity has become
very significant from the view point of cosmology. Though Einstein
ultimately abandoned it stating that it was a ``blunder'' in his
life but Tolman keeps it as a constant quantity in his field
equations even in 1939's famous work related to astrophysical
system. It is also, in favor of keeping $\Lambda$, argued by
Peebles and Ratra (2002) that like all energy, the zero-point
energy related to space has to contribute to the source term in
Einstein's gravitational field equations. However, it is being
gradually felt that the erstwhile cosmological constant $\Lambda$
is indeed a scalar variable dependent on time rather than a
constant as was being believed earlier. Recently, this variation
in cosmological constant is also observationally confirmed due to
the evidence of high redshift Type Ia supernova (Perlmutter et al.
1998; Riess et al. 1998) for a small decreasing value which is
$\leq 10^{-56} cm^{-2}$ at the present epoch. Obviously, once the
$\Lambda$ becomes a scalar its dependence need not be limited only
to time coordinate (as in cosmology). Since it enters in the field
equations as a variable, it must be dependent on space coordinates
as well. Therefore, in general, the $\Lambda$ is a scalar variable
depending on, either or both, space and time coordinates. It is
argued that just as in cosmology the dependence of $\Lambda$ on
time has been found to be of vital importance playing a
significant role now, its dependence on space coordinates is
equally important for astrophysical problems (Chen and Wu 1990;
Narlikar, Pecker and Vigier 1991; Ray and Ray 1993; Tiwari and Ray
1996).\\ With this view point, we consider here an anisotropic
charged static fluid sphere by introducing a scalar variable
$\Lambda$ dependent only on the radial coordinate $r$. The field
equations thus obtained, under certain mathematical assumptions,
yield a set of solutions which has another historical importance,
known in the name of Electromagnetic Mass Models (EMMM) in the
literature (Lorentz 1904; Hoffmann 1935; Feynman, Leighton and
Sands 1964; Tiwari, Rao and Kanakamedala 1986; Wilczek 1999). The
effective gravitational mass of these models depends on the
electromagnetic field alone, viz., the effective gravitational
mass vanishes when the charge density vanishes. Such models have
been studied by several authors (Ray and Ray 1993; Tiwari, Rao and
Kanakamedala 1984; Tiwari, Rao and Ray 1991; Gautreau 1985;
Gr{\o}n 1985, 1986a,b; Ponce de Leon 1987a,b, 1988; Tiwari and Ray
1991a,b, 1997; Ray, Ray and Tiwari 1993). All these EMMMs,
however, have been obtained under a special assumption $ \rho + p
= 0 $, where $ \rho $ is the matter-energy density and $ p $ is
the fluid pressure under the general condition that $\rho>0$ and
$p<0$. This type of equation of state, viz., $p=\gamma\rho$ with
$\gamma=-1$, implies that the matter distribution is in tension
and hence the matter is known, in the literature, as a `false
vacuum' or `degenerate vacuum' or `$\rho$-vacuum' (Davies 1984;
Blome and Priester 1984; Hogan 1984; Kaiser and Stebbins 1984). A
natural question arises whether there exists any EMMM where this
condition is violated, i.e., when $\rho + p \neq 0$. This is the
main motivation of the present investigation and here we have
shown that even for $\rho + p \neq 0 $ EMMM can be constructed.
However, the same question was addressed by Tiwari, Rao and Ray
(1991) and obtained EMMM in the isotropic and axially-symmetric
matter distribution for charged dust case only whereas Ray and
Bhadra (2004b) searched for solution to the problem by employing a
relation between the radial and tangential pressures as
$p_{\perp}= p_r + \alpha q^2 r^2$ where $\alpha$ is a non-zero
constant factor and $q$ is electric charge of the spherical system
of radius $r$. We shall consider in the present investigation
different relation and would like to observe that how this helps
us to find out several class of solutions related to EMMM.

\section{The Einstein-Maxwell Field Equations}

 Let us consider a spherically symmetric line element
\begin{equation}
ds^{2} = g_{ij}dx^{i}dx^{j} = e^{\nu(r)} dt^{2} - e^{\lambda(r)}
dr^{2} - r^{2} ( d \theta ^{2} + sin^{2} \theta d\phi^{2} ), \quad
(i,j = 0, 1, 2, 3)
\end{equation}
where $\nu$ and $\lambda$ are the metric potentials.\\
Now, the Einstein field equations for the case of charged
anisotropic source are
\begin{equation}
{G^{i}}_{j} \equiv {R^{i}}_{j} - \frac{1}{2}{{g^{i}}_{j}} R =
-\kappa [{{T^{i}}_{j}}^{(m)} + {{T^{i}}_{j}}^{(em)} +
{{T^{i}}_{j}}^{(vac)}] ,
\end{equation}
where ${T^{i}}_{j}^{(m)}$, ${T^{i}}_{j}^{(em)}$ and
${T^{i}}_{j}^{(vac)}$ are respectively the energy-momentum tensor
components for the anisotropic matter source, electromagnetic
field and vacuum. The explicit forms of these tensors are given by
\begin{equation}
{{T^{i}}_{j}}^{(m)} = (\rho + p_{\perp}) u^{i}u_{j} - p_{\perp} {
g^{i}}_{j} +(p_{\perp} -p_{r}){\eta}^{i}{\eta}_{j},
\end{equation}
\begin{equation}
{{T^{i}}_{j}}^{(em)} = -\frac{1}{4\pi} [ F_{jk}F^{ik} -
\frac{1}{4}
  {g^{i}}_{j}F_{kl} F^{kl}]
\end{equation}
and
\begin{equation}
{{T^{i}}_{j}}^{(vac)} = \frac{1}{8\pi} {g^{i}}_{j} \Lambda(r)
\end{equation}
with  $ u_{i}u^{i} = - {\eta}_{i}{\eta}^{i}=1$. The Maxwell
electromagnetic field equations are given by
\begin{equation}
{[{(- g)}^{1/2} F^{ij}],}_{j} = 4\pi J^{i}{(- g)}^{1/2}
\end{equation}
and
\begin{equation}
F_{[ij,k]} = 0,
\end{equation}
where the electromagnetic field tensor $F_{ij}$ is related to the
electromagnetic potentials through $ F_{ij} = A_{i,j} - A_{j,i} $
which, obviously, is equivalent to the equation (6.7). Further,
$u^{i}$ is the 4-velocity of a fluid element, $J^{i}$ is the
4-current satisfying $J^{i} = \sigma u^{i}$, where $\sigma$ is the
charge density, and $ \kappa = 8 \pi $ (in relativistic unit G = C
= 1). Here and in what follows a comma denotes the partial
derivative with respect to the coordinates (involving the
index).\\
 The Einstein-Maxwell field equations
(6.2) -- (6.7) corresponding to static anisotropic charged source
with cosmological variable, are then given by
\begin{equation}
e^{-\lambda} ( \lambda^{\prime}/r - 1/r^{2} ) + 1/r^{2} = 8 \pi
{{T^{0}}_{0}} = 8\pi \rho + E^{2} + \Lambda  ,
\end{equation}
\begin{equation}
e^{-\lambda} ( \nu^{\prime}/r + 1/r^{2} ) - 1/r^{2} = - 8 \pi
{{T^{1}}_{1}} = 8\pi {p}_{r} -E^{2} - \Lambda,
\end{equation}
\begin{equation}
e^{-\lambda} [ \nu^{{\prime}{\prime}}/2 + {\nu^{\prime}}^{2}/4 -
{\nu^{\prime} \lambda^{\prime}}/4 + (\nu^{\prime} -
\lambda^{\prime} )/ 2r] = - 8 \pi {{T^{2}}_{2}} = - 8 \pi
{{T^{3}}_{3}} = 8\pi p_{\perp} + E^{2} - \Lambda
\end{equation}
and
\begin{equation}
{(r^2 E)}^{\prime} = 4\pi r^2 \sigma e^{\lambda/2}.
\end{equation}
The equation (6.11) can equivalently, in terms of the electric
charge $q$, be expressed as
\begin{equation}
q(r) = r^2E(r) = \int_{0}^{r} 4 \pi r^2 \sigma e^{\lambda/2} dr
\end{equation}
 where $p_r$, $p_{\perp}$ and $E$ are the matter-energy
density, radial and tangential pressures and intensity of the
electric field respectively. Here prime denotes derivative with
respect to radial coordinate $r$ only.\\ The equation of
continuity, ${{{T^{i}}_{j}}};i = 0$ is given by
\begin{equation}
\frac{d}{dr}\left[p_r - \{E^2 + \Lambda(r)\} /{8 \pi}\right] + (
\rho + p_r ) {\nu^\prime}/2 = E^2/2 \pi r + 2( p_{\perp} - p_r)/r.
\end{equation}
Now, our spherically symmetric fluid distribution under
investigation is anisotropic in nature. This means that the radial
and tangential pressures are, in general, unequal so that the
simplest relation between them we assumed here as
\begin{equation}
p_{\perp} = n p_{r},\quad(n\neq 1).
\end{equation}
 Assuming further that the radial stress  $ {{T^{1}}_{1}}
= 0$ (Florides 1977; Kofinti 1985; Gr{\o}n 1986a; Ponce de Leon
1987b; Tiwari and Ray 1996) one gets
\begin{equation}
 \nu^{\prime}
= (e^{\lambda} - 1)/r
\end{equation}
and
\begin{equation}
p_r = [E^2 +\Lambda(r)]/8 \pi.
\end{equation}
Using equations (6.14) -- (6.16), in equation (6.13), we get
\begin{equation}
\rho + p_r = [(n + 1)E^2  + (n - 1)\Lambda(r)]/2 \pi(e^\lambda -
1).
\end{equation}
 Similarly, equations (6.8) and (6.9) yield
\begin{equation}
e^{-\lambda}( \nu^{\prime} + \lambda^{\prime}) = 8 \pi r( \rho +
p_r).
\end{equation}
Again, equation (6.8) through the equation (6.16), gives
\begin{equation}
e^{-\lambda} = 1 - 2m(r)/r,
\end{equation}
where m(r), called the effective gravitational mass, takes the
form
\begin{equation}
m(r) = M(r) + \mu(r) = 4 \pi \int_{0}^{r}[\rho + p_r] r^2 dr,
\end{equation}
the active gravitational mass of Schwarzschild type and the mass
equivalence of electromagnetic field, respectively, being defined
as
\begin{equation}
M(r) = 4 \pi \int_{0}^{r} \rho r^2 dr, \quad \mu(r) = 4 \pi
\int_{0}^{r} p_r r^2 dr.
\end{equation}
 Now, from the above equation (6.20) it is easily
observed that the condition $ \rho + p_r = 0 $, yields a flat
space-time through the equations (6.19) and (6.15) and has been
considered by Tiwari, Ray and Bhadra (2000) in another context.
Hence, the non-trivial solutions exits here for the case $ \rho +
p_r \neq 0 $ only.

\section{Solutions for the Static Charged Fluid
Spheres}

 Now we solve the equation (6.17) under the
constraint $ \rho + p_r \neq 0 $, assuming different mathematical
conditions. As the equation (6.17) is involved with two physical
parameters, $E$ and $\Lambda$, so unless we specify one parameter
in terms of other it is not possible to solve the equation (6.20).
Therefore, a plausible straightforward relation between these
parameters may be of the form $\Lambda(r)=\pm E(r)^2$. Physically
this means that vacuum energy has contribution from the
electrostatic field energy and proposal of this kind, in an
implicit way, is not at all unavailable in the literature (Ray and
Bhadra 2004b). Let us, therefore, assume the following two cases
$\Lambda(r) = E^2 - N\Lambda_{0}$ and $\Lambda(r) =- E^{2}+ N
{\Lambda}_{0}$ which will yield solutions with physically
interesting features as the analysis of the next section
demonstrates it clearly. However, in both the cases we have taken
the assumptions in a way so that the cosmological variable
$\Lambda$ does not vanish rather may be at most equal to
$\Lambda_{0}$, the erstwhile cosmological constant having a finite
non-zero value. Without considering $\Lambda_{0}$ we will have
$\Lambda = 0$ at the boundary $r = a$ which is a bit unphysical
and may create difficulties, such as to entropy like problems
(Beesham 1993). \\

 {\it Case I}\\
\noindent Let us assume that the cosmological constant is a
function of radial distance such that
\begin{equation}
\Lambda(r) = E^2 - N\Lambda_{0}
\end{equation}
where $N$ is a free parameter and $ \Lambda_{0} $ is the erstwhile
non-zero cosmological constant. With the help of equations (6.17)
and (6.22), the equation (6.20) takes the form
\begin{equation}
m(r) = 2 \int_{0}^{r}[2nE^2 - (n - 1)N \Lambda_{0}] r^2
dr/(e^{\lambda} - 1).
\end{equation}
To make equation (6.23) integrable we further assume that
\begin{equation}
E^2 = q^2/{r^4} = [k^2(e^{\lambda} - 1) (1 - R^2) + (n - 1) N
\Lambda_{0}] / 2n,
\end{equation}
where $ k $ is a constant and $ R = r/a $, $ a $ being the radius
of the sphere. This particular choice for the electric intensity
generates a model for charged sphere which is physically very
interesting as it is related to EMMM as will be seen later on.
Thus, the solution set is given by
\begin{equation}
e^{-\lambda(r)} = 1 - AR^2 (5 - 3R^2),
\end{equation}
\begin{equation}
e^{\nu(r)} = (1 - 2A)^{5/4} e^{\lambda(r)/4} exp[5B{tan^{-1}B(6R^2
- 5) - tan^{-1}B}/2 ],
\end{equation}
\begin{equation}
{p_{r}}(r) = {p_{\perp}}(r)/n = [k^2\{e^\lambda(r) - 1\}(1 - R^2)
- N \Lambda_{0}]/{8 \pi n}
\end{equation}
and
\begin{equation}
\rho(r) =  [k^2\{1 + 4n - e^\lambda(r)\}(1 - R^2) + N
\Lambda_{0}]/{8 \pi n},
\end{equation}
where
\begin{equation}
A = 4k^2a^2/15, \quad B = [A/(12 - 25A)]^{1/2}.
\end{equation}
Now, the exterior field of a spherically symmetric
static charged fluid distribution described by the metric (6.1) is
the unique Reissner-Nordstr{\"o}m solution given by
\begin{equation}
ds^{2} = \left[1 - \frac{2m}{r} + \frac{q^2}{r^2}\right] dt^{2} -
\left[1 - \frac{2m}{r} + \frac{q^2}{r^2}\right]^{-1} \ dr^{2} -
r^{2} ( d \theta ^{2} + sin^{2} \theta d\phi^{2} ).
\end{equation}
Then, application of the matching condition on the boundary $ r =
a $ yields the total effective gravitational mass in the following
form
\begin{equation}
m(a) =  \frac{4}{15}k^2a^3 + \frac{q(a)^2}{2a}.
\end{equation}
Hence, in terms of the total gravitational mass, the total
electric charge and the radius of the sphere, all the constants
$k$, $A$ and $B$ can be expressed as
\begin{equation}
k^2 = 15(2am - q^2)/8a^4,
\end{equation}
\begin{equation}
A = (2am - q^2)/2a^2
\end{equation}
and
\begin{equation}
B^2 = (2am - q^2)/[24a^2 - 25(2am - q^2)].
\end{equation}
 Comparing the equation (6.20) (and subsequently via the
equation (6.21)) with the above equation (6.31) we can easily
recognize the first term ($4k^2a^3/15$) in the right hand side as
the total Schwarzschild mass whereas the second term ($q^2/2a$) is
the total mass equivalence of the electromagnetic field. Now, $k$,
in the first term of equation (6.31), is implicitly related to the
charge $q$ which is evident from the equation (6.24). Therefore,
in principle, the gravitational mass $m$ expressed in equation
(6.31) is of purely electromagnetic in origin.\\

 {\it Case II}\\
 Here our choice is
\begin{equation}
\Lambda(r) =- E^{2}+ N {\Lambda}_{0}.
\end{equation}
Then, from the equations (6.17) and (6.35), the gravitational mass
(equation (6.20)) reduces to
\begin{equation}
m(r) = 2 \int_{0}^{r}[2E^2 + (n - 1)N \Lambda_{0}] r^2
dr/(e^{\lambda(r)} - 1).
\end{equation}
For the further assumption
\begin{equation}
E^2 = q^2/ r^4 = [k^2 (e^{\lambda(r)} - 1) (1 - R^2) - ( n - 1) N
\Lambda_{0}]/2,
\end{equation}
we obtained the solution set as
\begin{equation}
e^{-\lambda(r)} = 1 - AR^2 (5 - 3R^2),
\end{equation}
\begin{equation}
e^{\nu(r)} = (1 - 2A)^{5/4} e^{\lambda(r)/4} exp[5B{tan^{-1}B(6R^2
- 5) - tan^{-1}B}/2 ],
\end{equation}
\begin{equation}
p(r) = N\Lambda_{0}/8 \pi
\end{equation}
and
\begin{equation}
\rho(r) = [4k^2(1 - R^2) - N\Lambda_{0}]/8 \pi.
\end{equation}
We see that $ \lambda $ and $ \nu $ retain the same form as in the
case I and hence the total gravitational mass is also given by
the equation (6.31). Further, it is observed that the present case
automatically reduces to an isotropic one as the pressure $p$ does
not associated with the anisotropic factor $n$.

\section{Physical Properties of the Static Charged
Fluid Spheres}

 {\it Case I:~$\Lambda(r) = E^2 - N\Lambda_{0}$}\\

  {\it Subcase (1):}~The equation (6.24) indicates that for getting a direct dependence
of $k$ upon $q$ one can admit the following relaxation such that
(i) $N = 0$ when $\Lambda_0 \neq 0$ and $(n - 1) \neq 0$, (ii) $(n
- 1) = 0$ when $N \neq 0$ and $\Lambda_0 \neq 0$ and (iii)
$\Lambda_0 = 0$ when $ (n - 1) \neq 0$ and $N \neq 0$. The third
possibility seems to contradict the observational results related
to the Supernova type Ia where the cosmological constant is found
to be a non-zero positive value whereas the second one provides an
isotropic case. Then, suitably opting for $N = 0$ one obtains
\begin{equation}
q(r)^2 = k^2{r^4}(e^{\lambda(r)} - 1)(1 - R^2)/2n,
\end{equation}
\begin{equation}
{p_r}(r) = {p_{\perp}}(r)/n = k^2(e^{\lambda(r)} - 1)(1 - R^2)/{8
\pi n}
\end{equation}
and
\begin{equation}
\rho(r) = k^2\{1 + 4n - e^{\lambda(r)}\}(1 - R^2)/{8 \pi n}.
\end{equation}
Thus, for vanishing electric charge the gravitational mass in the
equation (6.31), including all the physical parameters (viz.,
pressures and density), vanishes and one obtains EMMM. \\

{\it Subcase (2):}~The central and the boundary pressures are
found to be equal here i.e., $p_r(0) = p_{\perp}(0)/n = -
N\Lambda_0/8\pi n$ and $p_r(a) = p_{\perp}(a)/n = -
N\Lambda_0/8\pi n$ respectively whereas the respective densities
are $(4nk^2 + N\Lambda_0)/8\pi n$ and $N\Lambda_0/8\pi n$. So, the
present model has a constant pressure throughout the sphere though
the density decreases from centre to boundary. For the value $N =
0$, however, we have zero pressure, both at the centre and
boundary. The density decreases from the non-zero central value
$k^2/2\pi$ and then smoothly decreases to zero at the boundary.
Thus, with $N = 0$ the present model goes to a physically
well-behaved static charged dust case.
\\
 {\it Subcase (3):}~In the above analysis for the
general value of $N$ we have seen that the fluid pressure is
altogether negative whereas the density is a positive quantity.
Now, it can be observed from the equations (6.27) and (6.28) that
$ \rho + p_r \neq 0 $ except at the boundary where it is equal to
$ p_r(a) = - \rho(a) $ such that $\rho(a) > 0$ and $p(a) < 0$ as
seen earlier. Otherwise, it will have a general value $ k^2 (1 -
R^2)/2\pi$ . The central value is then $ p_r(0) = -
\tilde{\rho}(0) $ where $\tilde{\rho}(0) = \rho(0) - k^2/2\pi$. As
the model demands for $\rho > 0$ and $p < 0$, so the condition to
be satisfied here is $\rho(0) >  k^2/2\pi$. The general condition
for the negative pressure and positive density is then $\rho > k^2
(1 - R^2)/2\pi$ for all $r \leq a$. These results are also true
for the sub-case $N = 0$.\\

\vspace{1.0cm}

 {\it Case II:~$\Lambda(r) = - E^{2}+ N {\Lambda}_{0}$}\\

{\it Subcase (1):}~Here for $N = 0$ the solution set, as obtained
in the equations (6.38), (6.40) and (6.41), reduces to
\begin{equation}
p(r) = 0
\end{equation}
and
\begin{equation}
\rho(r) = k^2(1 - R^2)/2 \pi,
\end{equation}
when the electric charge is given by
\begin{equation}
q(r)^2 = k^2 (e^{\lambda(r)} - 1) (1 - R^2)r^4/2.
\end{equation}
Thus, as in the previous case, for $q = 0$ we get $k = 0$ which in
turn makes mass, pressure and density to vanish and also the
space-time becomes flat. Thus, the model presented here is an
EMMM.\\ {\it Subcase (2):}~In the present case also the central
and the boundary pressures are equal with a value
$N\Lambda_0/8\pi$ and the respective densities are $(4k^2 -
N\Lambda_0)/8\pi $ and $ - N\Lambda_0/8\pi$. The pressures become
zero for $N = 0$,
 at the centre and boundary, and densities have the
 positive central value $k^2/2\pi$
whereas the boundary value is zero. Thus, we again get a
physically interesting charged dust case with $N = 0$ which now
corresponds to a case of isotropic fluid sphere. Here also the
behavior is regular and well defined.
\\
 {\it Subcase (3):}~In the present case also, by
virtue of equations (6.40) and (6.41), $\rho + p_r \neq 0 $ which
reads here as $\rho +  p_r =  k^2 (1 - R^2)/2\pi$. Here the
central value, at $r=0$, is $ \rho = - \tilde{p}$ where $\tilde{p}
= ( p - k^2/2\pi)$ and the boundary one is $ \rho = - p $. Due to
negative value of the density here the condition on the pressure
to be imposed is $ p > k^2/2\pi$. Thus, the present situation,
viz. {\it Case II}, clearly provides an EMMM even with a positive
pressure and therefore contradicts the comment made by Ivanov
(2002) that ``... electromagnetic mass models all seem to have
negative pressure.'' The same result, i.e. the positivity of
pressures are also available in some cases of the work done by Ray
and Das (2002) related to EMMM. However, the explanation given
here is valid for any positive value of $N$ and so the situation
could completely be opposite if one assigns on $N$ any negative
value. At this stage, we should not put any restriction on the
choice of the value of $N$. This is because, in general, for a
fluid sphere we should have $p \geq 0$ and $\rho \geq 0$ so that
the weak energy conditions are satisfied. But there are also some
special situations available within the spherical system
(particularly in the case of electron with the radius $\sim
10^{-16}$) where the energy condition is violated due to negative
energy density (Cooperstock and Rosen 1989; Bonnor and Cooperstock
1989; Ray and Bhadra 2004b). Thus, choosing the proper signatures
of $N$, we can have a class of models with diverse characters.

\vspace{1.0cm}

\section{Role of $\Lambda$: Previous and Present
Status}

 The cosmological constant was introduced by Einstein in
his field equation to obtain a static cosmological solution
because of the fact that due to gravitational pull everything will
collapse to a point and hence a un-wanting situation of
singularity will take place. However, he was not satisfied with
this new physical quantity as it seemed to violate Machian
principle which he tried to incorporate in the framework of his
General Relativity. He thus, ultimately rejected it mainly for two
reasons: (i) that the theoretical work of de Sitter showing that
the Einstein's field equations admitted a solution for empty
Universe and (ii) that the experimental discovery of expanding
Universe by Hubble.\\ As stated in the introduction, the concept
of cosmological constant has been revived recently in the case of
early Universe scenario and even in particle physics. It is
gradually being felt that $\Lambda$, the erstwhile cosmological
constant is available rather than a constant, as was being
believed earlier, varying with space or time or both (Tiwari and
Ray 1996; Ray, Ray and Tiwari 1993; Tiwari, Ray and Bhadra 2000).
Further, this varying $\Lambda$ may be positive or negative (by
imposing the condition that its value is not equal to zero). For
instance, according to Zel'dovich (1968) the effective
gravitational mass density of the polarized vacuum is negative.
Similarly, the equation of state $\rho + p = 0$, employed by
Tiwari, Rao and Kanakamedala (1984) to construct EMMM as a
solution of Einstein-Maxwell field equations, provides negative
pressure. It may be emphasized here that positive density has
significant, rather major role in inflationary cosmology whereas
negative density has influence on elementary particle models. The
gravitational mass inside the spherical charged body is negative
for $r<5a/4$, where $r$ is the radial coordinate and $a$ is the
radius of the sphere. It is argued by Gr{\o}n (1986a,b) that this
negative mass and the associated gravitational repulsion is due to
the strain of the vacuum because of vacuum polarization. He also
argued that if a vacuum has a vanishing energy, then its
gravitational mass will be negative and the observed expansion of
the universe may be explained as a result of repulsive
gravitation. Now, if we consider a negative $ \Lambda$ having a
repulsive nature as was considered by Einstein then this gets the
same status of negative pressure and also can be identified with
the Poincar{\'e} stress. This repulsive gravitation associated
with negative $\Lambda$ can also be explained as the source of
gravitational blue shift (Gr{\o}n, 1986a). On the contrary,
positive $\Lambda$ will be related to gravitational red shift. It
may also be pointed out that according to Ipser and Sikivie (1984)
domain walls are sources of repulsive gravitation and a spherical
domain wall will collapse. To overcome this situation the charged
``bubbles" with negative mass keep the wall static and hence in
equilibrium. In this regard, we may also add that $\Lambda$, via
repulsive gravitation, is related to domain walls and playing an
important physical role.\\ Very recent observations conducted by
the SCP and HZT (Perlmutter et al. 1998; Riess et al. 1998;
Filippenko 2001; Kastor and Traschen 2002) show that the present
value of $\Lambda$ is positive one and hence related to the
repulsive pressure. It is believed that the present state of
acceleration dominated Universe is due to the driven force of this
$\Lambda$. It is, therefore, to be noted that the negative
$\Lambda$ corresponds to a collapsing situation of the Universe
(Cardenas et al. 2002).

\vspace{1.5cm}

\section{Conclusions}

(i) In both the above cases {\it I} and {\it II}, it is possible
to show that EMMM can be obtained, in principle, using the
constraint $ \rho + p \neq 0 $. This particular point remained
unnoticed by Gr{\o}n (1986a,b) and Ponce de Leon (1987a,b) both.
\\ (ii) It can be noted that in terms of energy-momentum tensor of
the fluid the condition $ \rho + p = 0 $ implies $ {T^{1}}_{1} =
{T^{0}}_{0}$$^{43}$ whereas $ \rho + p \neq 0 $ constraint may be
expressed as $ { T^{1}}_{1} = 0 $ as we have adopted in the
present approach. It is also interesting to note that $ \rho + p =
0 $ and hence $ {T^{1}}_{1} = {T^{0}}_{0}$ can be expressed in
terms of the metric tensors (vide equation (6.1)) as $g_{00}g_{11}
= -1$. A coordinate-independent statement of this relation is
obtained by Tiwari, Rao and Kanakamedala 1984) by using the eigen
values of the Einstein tensor ${G^{i}}_{j}$.\\ (iii) We would like
to mention here that the solutions obtained by Gr{\o}n (1986a,b)
and Ponce de Leon (1987a,b) represent a neutral system, viz.,
though the net charge is not zero but the charge on the surface of
the spherical system vanishes. The models of the present paper, in
general, do not correspond to this situation because of the fact
that the electric field and hence cosmological constant does not
vanish at the boundary. In both the {\it Case I} and {\it Case
II}, the values of electric field, respectively, are $(n - 1)N
\Lambda_{0}/2n$ and $-( n - 1)N \Lambda_{0}/2$ whereas those for
cosmological parameters are $-(n+1)N\Lambda_{0}/2n$ and
$(n+1)N\Lambda_{0}/2$. Therefore, the present solutions correspond
to a charged fluid sphere. Of course, for $N = 0$, like Gr{\o}n
(1986a,b) and Ponce de Leon (1987a,b), we have neutral spheres
(equation (6.42) of the {\it Case I} and equation (6.47) of the
{\it Case II}). Thus, we have a class of solutions related to
charged as well as neutral systems depending on the values of $N$.

\vspace{17.0cm}

{\it The contents of this chapter has communicated to journal for
publication.}

\pagebreak

\chapter{Conclusions}

\vspace{1.0cm}

\hrule \vspace{0.5cm}
 \begin{quote}
 {\it ``So we come back again to the original idea of
Lorentz -- may be all the mass of an electron is purely
electromagnetic, may be the whole 0.511 Mev is due to
electrodynamics."}\\ -- Feynman et al. (1964)
\end{quote}
\hrule

\vspace{1.5cm}

Electromagnetic mass models which are the sources of purely
electromagnetic origin ``have not only heuristic flavor associated
with the conjecture of Lorentz but even a physics having
unconventional yet novel features characterizing their own
contributions independent of the rest of the physics" (Tiwari
2001). This is, as Tiwari (2001) guess ``may be due to the subtle
nature of the mass of the source (being dependent on the
electromagnetic field alone)". Therefore, in our whole attempt we
have tried to explore ``the subtle nature of the mass of the
source". However, to do this under the general relativistic
framework, we have considered Einstein field equations in its
general form, i.e., with cosmological constant $\Lambda$ which
also acts as a source term to the energy-momentum tensor. If we
consider that $\Lambda$ has a variable structure which is
dependent on the radial coordinate of the spherical distribution,
viz., $\Lambda= \Lambda(r)$ then it can be shown that $\Lambda$ is
related to pressure and matter energy density. Hence it
contributes to the effective gravitational mass of the system.\\
It is seen that equation of state has an important role in
connection to electromagnetic mass model. Therefore, at first we
have obtained electromagnetic mass model under the condition
$\rho+p=0$. However, later on it is shown that electromagnetic
mass model can also be obtained  by using more general condition
$\rho+p \neq 0$.\\ The model considered in our work, in general,
corresponds to a charged sphere with cosmological parameter in
such a way that it does not vanish at the boundary. The idea
behind is that the cosmological parameter is related to the zero
point vacuum energy it should have some finite non-zero value even
at the surface of the bounding system. For this type of spherical
system we can have a class of solutions related to charged as well
as neutral configurations.\\ It can be shown that these models
have positive energy densities everywhere. Their corresponding
radii are always much larger than $10^{-16}$ cm. Furthermore, as
the radii of these models shrink to zero, their total
gravitational mass becomes infinite. It have been shown by Bonnor
and Cooperstock (1989) that an electron must have a negative
energy distribution (at least for some values of the radial
coordinate). In this connection we have shown that the
cosmological parameter $\Lambda$ has a definite role on the energy
density of micro particle, like electron. At an early epoch of the
universe when the numerical value of negative $\Lambda$ was higher
than that of the energy density $\rho$, the later quantity became
a positive one. In the case of decreasing negative value of
$\Lambda$ there was a smooth crossover from positive energy
density to a negative energy density.\\ So far we have referred
electron to be a spherically symmetric distribution of matter
deprived of spin and magnetic moment. As an alternative way both
Bonnor and Cooperstock (1989) as well as Herrera and Varela (1994)
suggest that both spin and magnetic moment can be introduced at
classical level through the Kerr-Newman metric. However in this
context it is to be mentioned here that the Kerr-Newman metric
cannot be valid for distance scales of the radius of a subatomic
particle. We, therefore, thought that the problem can be tackled
in the frame work of Einstein-Cartan theory where torsion and spin
are inherently present. In this case, the only way is to take the
spin to be the `intrinsic angular momentum' that is the spin of
quantum mechanical origin. In our work considering the spins of
all the individual particles are assumed to be oriented along the
radial axis of the spherical systems we have obtained some
interesting solutions with physical validity. However, though our
present approach via Einstein-Cartan theory to inject spin may be
interesting it, at once, demands some alternative means to provide
spin and magnetic moment. This may be possible through
Dirac-Maxwell theory where spin and magnetic moment are naturally
incorporated through the Dirac spin. We would like to pursue this
problem in future investigations.\\ Another important point we
would like to mention here that in all the previous investigations
we have studied electromagnetic mass models in 4-dimensional
Einstein-Maxwell space-times only. Therefore, one can ask whether
electromagnetic mass models also can exist in higher dimensional
theory of General Relativity. We have presented a model which
corresponds to spherically symmetric gravitational sources of
purely electromagnetic origin in the space-time of $(n+2)$
dimensional theory of general relativity.\\ We have also taken up
the problem of anisotropic fluid sphere as studied earlier in a
different view point. By expressing $\Lambda$ in terms of electric
field strength $E$ we have explored some possibilities to
construct electromagnetic mass models using the constraint $ \rho
+ p \neq 0 $. We would like to mention here that unlike the
solutions of Gr{\o}n (1986a,b) and Ponce de Leon (1987a,b) in the
present investigation, in general, the electric field (and hence
the cosmological constant) does not vanish at the boundary.
However, it is shown that the class of solutions obtained here are
related to charged as well as neutral systems of Gr{\o}n (1986a,b)
and Ponce de Leon (1987a,b) depending on the values of the
parameter $N$.\\ It is to be mentioned here that other than
Dirac-Maxwell theory where spin and magnetic moment are naturally
incorporated through the Dirac spin, some other possibilities are
awaiting to be investigated under the scheme of electromagnetic
mass models. One of such possibilities is to study the
relationship between the structures of soliton which have been
identified with the electromagnetic field to that of electrons
which are also identified with the electromagnetic field (Tiwari
2001). This can be done by the use of Zakharov-Belinsky method to
solve the Einstein-Maxwell equations. Another possibility is to
conjecture that Weyl line-mass solutions and cosmic strings are
identical entities, because it has been shown by Linet (1985) and
Hiscock (1985) that the Weyl line-mass solutions can be identified
with the cosmic strings. On the other hand Weyl line-mass
solutions have been identified with the electromagnetic mass
models (Tiwari et al. 1991).

\pagebreak

\addcontentsline{toc}{chapter}{List of Publications}

\begin{center}
{\large\bf List of Publications}\\
\end{center}
\hrule

\vspace{0.5cm}

 1. R. N. Tiwari, Saibal Ray and {\bf Sumana Bhadra}, ``Relativistic Electromagnetic
 Mass Models with Cosmological Variable ${\Lambda}$ in Spherically Symmetric
Anisotropic Source" \\ {\it Indian Journal of pure and applied
Mathematics} (2000) {\bf 31} 1017.\\

2. Saibal Ray and {\bf Sumana Bhadra}, ``Classical Electron Model
with Negative Energy Density in Einstein-Cartan Theory of Gragitation" \\
{\it International
Journal of Modern Physics D} (2004) {\bf 13} 555.\\

3. Saibal Ray and {\bf Sumana Bhadra}, ``Energy Density in General Relativity:
a Possible Role for Cosmological Constant" \\ {\it Physics Letters A} (2004) {\bf 322}
150.\\

 4. Saibal Ray, {\bf Sumana Bhadra} and G. Mohanty, ``Relativistic Electromagnetic
 Mass Models: Charged Dust Distribution in Higher Dimensions" \\ {\it Astrophysics
 and Space Science} (2006) {\bf 302} 153.\\

5. Saibal Ray, {\bf Sumana Bhadra} and G. Mohanty, ``Relativistic Anisotropic
Charged Fluid Spheres with Varying Cosmological Constant" \\
{\it Communicated to journal}. \\

\pagebreak

\addcontentsline{toc}{chapter}{References}

\begin{center}
{\large\bf References}
\end{center}

\vspace{1.5cm}

\noindent {\Large\bf A}bdel-Rahman, A.- M. M.
(1990) {\it Gen. Rel. Grav.} {\bf 22} 655.\\

\noindent Abraham, M. (1902) {\it Dynamik  dee Elektrons},
(Gottinger Nachr., 20-41).\\

\noindent Abraham, M. (1905) {\it Theorie der Elektrizital : II,
Electro-magnetische Theorie der Strahlung} (Teubner, Leipzig).\\

\noindent Arkuszewski, W., Kopczynski, W. and Ponomariev,
              V. N. (1975) {\it Comm. Math.  Phys.} {\bf 45} 183.\\

\noindent {\Large\bf B}anerjee, A., Bhui, B. K. and Chatterjee, S.
(1990) {\it Astron. Astrophys.} {\bf 232} 305.\\

\noindent Bayin, S. S. (1978) {\it Phys. Rev. D} {\bf {18}}
2745.\\

\noindent Baylin, M. and Eimerl, D. (1972) {\it {Phys. Rev. D}}
{\bf {5}} 1897.\\

\noindent Beesham, A. (1993) {\it Phys. Rev. D} {\bf 48} 3539.\\

\noindent Berman, M. S. (1990a) {\it Int. J. Theor. Phys.} {\bf
29} 567.\\

\noindent Berman, M. S. (1990b) {\it Int. J. Theor. Phys.} {\bf
29} 1419.\\

\noindent Berman, M. S. (1991a) {\it Gen. Rel. Grav.} {\bf 23}
465.\\

\noindent Berman, M. S. (1991b) {\it Phys. Rev. D}
             {\bf 43} 1075.\\

\noindent Berman, M. S. and  Som, M. M. (1990) {\it Int. J. Theor.
Phys.} {\bf 29} 1411.\\

\noindent Berman, M. S., Som, M. M. and  Gomide, F. M. (1989) {\it
Gen. Rel. Grav.} {\bf 21} 287.\\

\noindent Bertolami, O. (1986a) {\it Nuo. Cim. B} {\bf 93} 36.\\

\noindent Bertolami, O. (1986b) {\it Fortschr. Phys.} {\bf 34}
829.\\

\noindent Bhui, B. K. and Chatterjee, S.
(1990) {\it Astron. Astrophys.} {\bf 232} 305.\\

\noindent Blinder, S. M. (2001) {\it Rept. Math. Phys.} {\bf 47}
279.\\

\noindent Blome, J. J. and Priester,  W. (1984) {\it
Naturwissenshaften} {\bf 71} 528.\\

\noindent Bonnor, W. B. and Cooperstock, F. I. (1989) {\it Phys.
Lett. A} {\bf 139} 442.\\

\noindent Bohun, C. S. and Cooperstock, F. I. (1999) {\it Phys.
Rev. A} {\bf 60} 4291.\\

\noindent Bopp, F. (1940)  {\it Ann. Phys., Lpz.} {\bf 38} 345.\\

\noindent Bopp, F. (1943)  {\it Ann. Phys., Lpz.} {\bf 42} 573.\\

\noindent Born, M. (1962) {\it Einstein's Theory of Relativity}
Chap. V (Dover, New York, p~372).\\

\noindent Born, M. and Infeld, L. (1934) {\it Proc. R. Soc.
London} {\bf A144} 425.\\

\noindent Bucherer, A. H. (1908) {\it Phys. Z.} {\bf 9} 755.\\

\noindent {\Large\bf C}ardenas, R. et al. (2002)
astro-ph/0206315.\\

\noindent Carvalho, J. C., Lima,  J. A. S. and Waga, I. (1992)
{\it Phys. Rev. D} {\bf 46} 2404.\\

\noindent Casimir, H. B. G. (1948) {\it Proc. K. Ned. Acad. Wet.}
{\bf 51} 635.\\

\noindent Chatterjee, S. and Bhui, B. K. (1990) {\it Astrophys.
Space Sci.} {\bf 167} 61.\\

\noindent Chen, W. and Wu, Y. S. (1990) {\it Phys. Rev. D} {\bf
41} 695. \\

\noindent Chodos, A. and Detweiler, S. (1980) {\it Phys. Rev. D}
{\bf 21} 2167.\\

\noindent Cohen, J. M. and Cohen, M. D. (1969) {\it Nuovo Cimento
B} {\bf 60} 241.\\

\noindent Cohen, J. M. and Gautreau, R. (1979) {\it Phys. Rev. D} {\bf 19} 2273.\\

\noindent Cooperstock, F. I. and de la Cruz, V. (1978) {\it Gen.
Rel. Grav.} {\bf 9} 835.\\

\noindent Cooperstock F. I. and Rosen N. (1989) {\it Int. J.
Theor. Phys.} {\bf 28} 423.\\

\noindent {\Large\bf D}avies, C. W. (1984) {\it Phys. Rev. D} {\bf
30} 737.\\

\noindent de la Cruz, V. and Israel, W. (1967) {\it Nuo. Cim.}
{\bf 51} 744.\\

\noindent de Sabbata, V. and Sivaram, C. (1994) {\it Spin and
Torsion in Gravitation} (World Scientific, Singapore, Chap.
VII).\\

\noindent De, U. K. and Raychaudhuri, A. K. (1968) {\it Proc. Roy.
Soc. London A} {\bf 303} 97.\\

\noindent Devitt, J. and Florides, P. S. (1989) {\it Gen. Rel.
Grav.} {\bf 21} 585.\\

\noindent Dirac, P. A. M. (1928) {\it Proc. R. Soc. London} {\bf
A118} 351.\\

\noindent {\Large\bf E}instein, A. (1905) {\it Ann. Physik}  {\bf
17} 891.\\

\noindent Einstein, A. (1919) {\it Sitz. Preuss. Akad. Wiss.} {\bf
349} (Reprinted in The Principle of Relativity, Dover, INC., N.Y.,
1952, p~190-198).\\

\noindent {\Large\bf F}ermi, E. (1922) {\it Z. Physik} {\bf 24}
340.\\

\noindent Feynman, R. P., Leighton, R. R. and Sands, M. (1964)
{\it The Feynman Lectures on Physics} (Addison-Wesley, Palo Alto,
Vol.II, Chap. 28). \\

\noindent Filippenko, A. V. (2001) astro-ph/0109399.\\

\noindent Florides, P. S. (1962) {\it Proc. Camb. Phil. Soc.} {\bf
58} 102.\\

\noindent Florides, P. S. (1977) {\it Nuovo Cimento} {\bf 42A}
343.\\

\noindent Florides, P. S. (1983) {\it Phys. A: Math. Gen.} {\bf
16} 1419.\\

\noindent Freese, K. et al. (1987) {\it Nucl. Phys. B} {\bf 287}
797.\\

\noindent Fukui, T. (1987) {\it Gen. Rel. Grav.} {\bf 19} 43.\\

\noindent {\Large\bf G}autreau, R. (1985) {\it Phys. Rev. D} {\bf
31} 1860.\\

\noindent Gegenberg, J. D. and Das, A. (1985) {\it Phys. Lett. A}
{\bf 112} 427.\\

\noindent Georgy, H., Quinn, H. and Weinberg, S. (1974) {\it Phys.
Rev. Lett.} {\bf 33} 451.\\

\noindent Gliner, E. B. (1966) {\it Sov. Phys. JETP} {\bf 22}
378.\\

\noindent Gr{\o}n, {\O}. (1985) {\it Phys. Rev. D} {\bf 31}
2129.\\

\noindent Gr{\o}n, {\O}. (1986a) {\it Am. J. Phys.} {\bf 54} 46.\\

\noindent Gr{\o}n, {\O}. (1986b) {\it Gen. Rel.
            Grav.} {\bf 18} 591.\\

\noindent Gunn, J. and Tinslay, B. M. (1975) {\it Nature} {\bf
257} 454.\\

\noindent {\Large\bf H}awking, S. W. and Ellis, G. F. R. (1973)
{\it The Large Scale Structure of Space-time} (Cambridge
University Press).\\

\noindent Heaviside, O. (1989) {\it Phil. Mag.}  {\bf 27} 324.\\

\noindent Hehl, F. W., von der Heyde, P. and Kerlick, G. D. (1974)
{\it Phys. Rev. D} {\bf 10} 1066.\\

\noindent Herrera, L. and  Varela, V. (1994) {\it Phys. Lett. A}
{\bf 189} 11.\\

\noindent Hiscock, W. A. (1985) {\it Phys. Rev. D} {\bf 31} 3288.
\\

\noindent Hoffmann B. (1935a) {\it Proc. R. Soc. London} {\bf
A148} 353.\\

\noindent Hoffmann, B. (1935b) {\it Phys. Rev. D} {\bf 47} 877.\\

\noindent Hogan, C. (1984) {\it Nature} {\bf 310} 365.\\

\noindent {\Large\bf I}ban{\`e}z, J. and Verdaguer, E. (1986) {\it
Phys. Rev D} {\bf 34} 1202.\\

\noindent Ishihara, H. (1984) {\it Prog. Theor. Phys.} {\bf 72}
376.\\

\noindent Ipser, J. and Sikivie, P. (1984) {\it Phys. Rev. D} {\bf
30} 712.\\

\noindent Ivanov, B. V. (2002) {\it Phys. Rev. D} {\bf 65}
104001.\\

\noindent {\Large\bf K}aiser, N. and Stebbins, A. (1984) {\it
Nature} {\bf 310} 391.\\

\noindent Kalligas, D., Wesson, P. and Everitt, C. W. F. (1992)
{\it Gen.Rel. Grav.} {\bf 24} 351.\\

\noindent Kaluza, T. (1921) {\it Sitz. Preuss. Akad. Wiss.} 966.\\

\noindent Kastor, D. and Traschen, J. (2002) {\it Class. Quantum
Grav.} {\bf 19} 5901.\\

\noindent Katz, J. and Horwitz, G. (1971) {\it Nuovo Cimento} {\bf
3B} 245-261.\\

\noindent Kaufmann, W. (1901a) {\it Phys. Z.} {\bf 3} 9.\\

\noindent Kaufmann, W. (1901b) {\it Phys. Z.} {\bf 2} 602.\\

\noindent Kaufmann, W. (1901c) {\it Gottinger Nachr.} 143-155.\\

\noindent Kaufmann, W. (1902a) {\it Gottinger Nachr.} 291-303.\\

\noindent Kaufmann, W. (1902b) {\it Phys. Z.} {\bf 4} 54.\\

\noindent Khadekar, G. S., Butey, B. P. and Shobhane, P. D. (2001)
{\it Jr. Ind. Math. Soc.} {\bf 68} 33.\\

\noindent Klein, O. (1926a) {\it Nature} {\bf 118} 516.\\

\noindent Klein, O. (1926b) {\it Z. Phys.} {\bf 37} 895.\\

\noindent Klein, O. (1928) {\it Z. Phys.} {\bf 46} 188.\\

\noindent Kofinti, N. K. (1985) {\it Gen. Rel. Grav.} {\bf 17}
245.\\

\noindent Kyle, C. F. and Martin, A. W. (1967) {\it Nuo. Cim. B}
{\bf 50} 583.\\

\noindent {\Large\bf L}au, Y. K. (1985) {\it Aust. J. Phys.} {\bf
29} 339.\\

\noindent Linde, A. D. (1984) {\it Rep. Prog. Phys.} {\bf 47}
925.\\

\noindent Linet, B. (1985) {\it  Gen. Rel. Grav.} {\bf 17} 1109.\\

\noindent Lisi, E (1995) {\it Journal of Physics A} {\bf 28}
5385.\\

\noindent L\'{o}pez, C.A. (1984) {\it Phys. Rev. D} {\bf 30}
313.\\

\noindent L\'{o}pez, C.A. (1986) {\it Phys. Rev. D} {\bf 33}
2489.\\

\noindent L\'{o}pez, C. A. (1988) {\it Phys. Rev. D} {\bf 38}
3662.\\

\noindent L\'{o}pez, C. A. (1992) {\it Gen. Rel. Grav.} {\bf 24}
285.\\

\noindent Lorentz, H. A. (1892) {\it Arch. Weerl.} {\bf 25} 363
[Reprinted in `Collected Papers' {\bf 4} 219].\\

\noindent Lorentz, H. A. (1895) {\it E. J. Brill., Leiden},
[Reprinted in `Collected Papers', {\bf 5} 1].\\

\noindent Lorentz, H. A. (1904) {\it Proc. Acad. Sci., Amsterdam}
6 (Reprinted in The Principle of Relativity, Dover, INC., 1952,
p~24).\\

\noindent Lorenz-Petzold, D. (1984) {\it Phys. Lett. B} {\bf 149}
79.\\

\noindent {\Large\bf M}ann, R. and Morris, M. (1993) {\it Phys.
Lett. A} {\bf 181} 443. \\

\noindent Martins, C. J. A. (2002) astro-ph/0205504. \\

\noindent Mehra, J. (Ed.) (1973): {\it The Physicist's conception of
Nature} ( D. Reidel Publishing Co., Dordrecht, Holland, p~331 -
369).\\

\noindent Mie, G. (1912a) {\it Ann. Phys., Lpz.}, {\bf 37} 511.\\

\noindent Mie, G. (1912b) {\it Ann. Phys., Lpz.}, {\bf 39} 1.\\

\noindent {\Large\bf N}arlikar, J. V., Pecker, J. -C  and Vigier,
J. -P (1991) {\it J. Astrophys. Astr.} {\bf 12} 7.\\

\noindent Newman, E., Couch, F., Chinapared, K., Exton, A.,
Prakash, A. and Torrence, R. (1965). {\it{J. Math. Phys.}} {\bf 6}
918.\\

\noindent Ng, Y. J. (1992) {\it Int. J. Theor. Phys.} {\bf 1}
154.\\

\noindent Novello, M. (1976) {\it Phys. Lett. A} {\bf 59} 105.\\

\noindent {\Large\bf {\"O}}zer, M. and Taha, M. O. (1986) {\it
Phys. Lett. B} {\bf 171} 363.\\

\noindent {\Large\bf P}ant, D. N. and Sah, A. (1979) {\it J.
Maths. Phys.} {\bf 20} 2537.
\\

\noindent Peebles, P. J. E. and Ratra, B. (1988) {\it Astrophys.
J.} {\bf 325} L17.\\

 \noindent
Peebles, P. J. E. and Ratra, B. (2002) {\it Rev. Mod. Phys.} {\bf
75} 559.\\

 \noindent
Perlmutter, S. et al. (1998) {\it Nature} {\bf 391} 51.\\

\noindent Poincar{\'e}, H. (1905) {\it C. R. Acad. Sci.} {\bf 140}
1504.\\

\noindent Poincar{\'e}, H. (1906) {\it R. C. Mat. Palermo} {\bf
21} 129.\\

\noindent Ponce de Leon, J. (1987a) {\it J. Math. Phys.} {\bf 28}
410.\\

\noindent Ponce de Leon, J. (1987b) {\it Gen. Rel. Grav.} {\bf 19}
797.\\

\noindent Ponce de Leon, J. (1988) {\it J. Maths. Phys.} {\bf 29}
197.\\

\noindent Prasanna, A. R. (1975) {\it Phys. Rev. D} {\bf 11}
2076.\\

\noindent {\Large\bf Q}uigg, C. (1983) {\it Gauge theories of the
strong, weak and electromagnetic interactions} (Benjamin, New
York, p~3).\\

\noindent {\Large\bf R}ay, S. (2007) {\it Apeiron} {\bf 14} 153 [arXiv: physics/0411127].\\

\noindent Raychaudhuri, A. K. (1979) {\it Theoretical Cosmology}
(Clarendon Press, Oxford, Chap.~10).\\

\noindent  Ray S. and Bhadra S. (2004a) {\it Int. J. Mod. Phys. D}
{\bf 13} 555.\\

\noindent  Ray S. and Bhadra S. (2004b) {\it Phys. Lett. A} {\bf
322} 150.\\

\noindent Ray S. and Das B. (2002) {\it Astrophys. Space Sci.}
{\bf 282} 635.\\

\noindent Ray S. and Das B. (2004) {\it Mon. Not. R. Astron. Soc.}
{\bf 349} 1331.\\

\noindent Ray S. and Das B. (2007a) {\it Grav. Cosmol.} {\bf 13} 224 arXiv: astro-ph/0409527].\\

\noindent Ray S. and Das B. (2007b) to appear in {\it Int. J. Mod. Phys. D} [arXiv: astro-ph/0705.2444].\\

\noindent Ray, S. and Ray, D. (1993) {\it Astrophys. Space Sci.}
{\bf 203} 211. \\

\noindent Ray, S., Ray, D. and Tiwari, R. N. (1993) {\it
Astrophys. Space Sci.} {\bf 199} 333.\\

\noindent Reuter, M. and Wetterich, C. (1987) {\it Phys. Lett. B}
{\bf 188} 38.\\

\noindent Riess, A. G. et al. (1998) {\it Astron. J.} {\bf 116}
1009.\\

\noindent {\Large\bf S}akharov, A. D. (1968) {\it Doklady Akad.
Nauk. SSSR} {\bf 177} 70 (translated: {\it Soviet Phys. Doklady}
{\bf 12}).\\

\noindent Schwarz, J. H. (1985) {\it Superstings} (World
Scientific, Singapore).\\

\noindent Searle, G. F. C. (1897) {\it Phil. Mag.} {\bf 44} 329.\\

\noindent Sistero, R. F. (1991) {\it Gen. Rel. Grav.} {\bf 23}
1265.\\

\noindent Som, M. M. and Bedran, M. L. (1981) {\it Phys. Rev. D}
{\bf 24} 2561.\\

\noindent {\Large\bf T}homson, J. J. (1881) {\it Phil. Mag.} {\bf
11} 229.\\

\noindent Thomson, J. J. (1897) {\it Phil. Mag.} {\bf 44} 298.\\

\noindent Tiwari, R. N. (2001) {\it Proceedings of International
Conference on Mathematical Physics} (vol. I, p. 27, Nagpur
University, India).\\

\noindent Tiwari, R. N., Rao, J. R. and Kanakamedala, R. R. (1984)
{\it Phys. Rev.D} {\bf 30} 489.\\

\noindent Tiwari, R. N., Rao, J. R. and Kanakamedala, R. R. (1986)
{\it Phys. Rev. D} {\bf 34} 1205.\\

\noindent  Tiwari, R. N., Rao, J. R. and Ray, S. (1991) {\it
Astrophys. Space Sci.} {\bf 178} 119.\\

\noindent Tiwari, R. N. and Ray, S. (1991a) {\it Astrophys. Space
Sci.} {\bf 180} 143.\\

\noindent Tiwari, R. N. and Ray, S. (1991b) {\it Astrophys. Space
Sci.} {\bf 182} 105.\\

\noindent Tiwari, R. N. and Ray, S. (1996) {\it Ind. J. Pure Appl.
Math.} {\bf 27} 907.\\

\noindent Tiwari, R. N. and Ray, S. (1997) {\it Gen. Rel. Grav.}
{\bf 29} 683. \\

\noindent Tiwari, R. N., Ray, S. and Bhadra, S. (2000) {\it Ind.
J. Pure Appl. Math.} {\bf 31} 1017.\\

\noindent Tolman, R. C. (1939) {\it Phys. Rev.} {\bf 55} 364.\\

\noindent {\Large\bf V}isser, M. (1989) {\it Phys. Lett. A} {\bf
139} 99.\\

\noindent {\Large\bf W}einberg, S. (1986) {\it Strings and Superstrings}
(World Scientific, Singapore).\\

\noindent Wienberg, S. (1989) {\it Rev. Mod. Phys.} {\bf 61} 1.\\

\noindent Wesson P. S. (1983) {\it Astron. Astrophys.} {\bf 119}
145.\\

\noindent Weyl, H. (1918a) {\it Sitz. Press. Akad. Wiss.} 465.\\

\noindent Weyl, H. (1918b) {\it Math. Z.} {\bf 2} 384.\\

\noindent Weyl, H. (1919) {\it Ann. Phys.} {\bf 54} 117.\\

\noindent Wilczek, F. (1998) {\it Nature} {\bf 394} 13.\\

\noindent Wilczek, F. (1999) {\it Phys. Today}  p. 11. \\

\noindent {\Large \bf Z}el'dovich, Ya. B.  (1967) {\it JETP
letters} {\bf 6} 316; Sov. Phys. Uspekhi 11 (1968) 381.\\

\end{document}